\documentclass[aps,prb,twocolumn,superscriptaddress,showpacs]{revtex4}
\usepackage{graphicx}
\usepackage{dcolumn}
\usepackage[T1]{fontenc} 
\usepackage{multirow}
\usepackage{amsmath} 
\begin{document}

\title{Strong Correlations, Strong Coupling and $s$-wave Superconductivity in Hole-doped BaFe$_{2}$As$_{2} $ Single Crystals}
\author{F. Hardy}
\email[]{frederic.hardy@kit.edu}
\affiliation{Karlsruher Institut f\"ur Technologie, Institut f\"ur Festk\"orperphysik, 76021 Karlsruhe, Germany}
\author{A. E. B\"ohmer}
\affiliation{Karlsruher Institut f\"ur Technologie, Institut f\"ur Festk\"orperphysik, 76021 Karlsruhe, Germany}
\author{L. de' Medici}
\affiliation{European Synchrotron Radiation Facility, BP 220, F-38043 Grenoble Cedex 9, France}
\author{M. Capone}
\author{G. Giovannetti}
\affiliation{CNR-IOM-Democritos National Simulation Centre and International School for Advanced Studies (SISSA), Via Bonomea 265, I-34136, Trieste, Italy}
\author{R. Eder}
\affiliation{Karlsruher Institut f\"ur Technologie, Institut f\"ur Festk\"orperphysik, 76021 Karlsruhe, Germany}
\author{L. Wang}
\affiliation{Karlsruher Institut f\"ur Technologie, Institut f\"ur Festk\"orperphysik, 76021 Karlsruhe, Germany}
\author{M. He}
\affiliation{Karlsruher Institut f\"ur Technologie, Institut f\"ur Festk\"orperphysik, 76021 Karlsruhe, Germany}
\author{T. Wolf}
\affiliation{Karlsruher Institut f\"ur Technologie, Institut f\"ur Festk\"orperphysik, 76021 Karlsruhe, Germany}
\author{P. Schweiss}
\affiliation{Karlsruher Institut f\"ur Technologie, Institut f\"ur Festk\"orperphysik, 76021 Karlsruhe, Germany}
\author{R. Heid}
\author{A. Herbig}
\affiliation{Karlsruher Institut f\"ur Technologie, Institut f\"ur Festk\"orperphysik, 76021 Karlsruhe, Germany}
\author{P. Adelmann}
\affiliation{Karlsruher Institut f\"ur Technologie, Institut f\"ur Festk\"orperphysik, 76021 Karlsruhe, Germany}
\author{R. A. Fisher}
\affiliation{Lawrence Berkeley National Laboratory, Berkeley, CA 94720, USA}
\author{C. Meingast}
\affiliation{Karlsruher Institut f\"ur Technologie, Institut f\"ur Festk\"orperphysik, 76021 Karlsruhe, Germany}
\begin{abstract}
We present a comprehensive study of the low-temperature heat capacity and thermal expansion of single crystals of the hole-doped Ba$_{1-x}$K$_{x}$Fe$_{2}$As$_{2}$ series ($0<x<1$) and the end-members RbFe$_{2}$As$_{2}$ and CsFe$_{2}$As$_{2}$.  A large increase of the Sommerfeld coefficient $\gamma_{n}$ is observed with both decreasing band filling and isovalent substitution (K, Rb, Cs) revealing a strong enhancement of electron correlations and the possible proximity of these materials to a Mott insulator. This trend is well reproduced theoretically by our Density-Functional Theory + Slave-Spin (DFT+SS) calculations, confirming that 122-iron pnictides are effectively Hund metals, in which sizable Hund's coupling and orbital selectivity are the key ingredients for tuning correlations. We also find direct evidence for the existence of a coherence-incoherence crossover between a low-temperature heavy Fermi liquid and a highly incoherent high-temperature regime similar to heavy fermion systems. In the superconducting state, clear signatures of multiband superconductivity are observed with no evidence for nodes in the energy gaps, ruling out the existence of a doping-induced change of symmetry (from $s$ to $d$-wave). We argue that the disappearance of the electron band in the range $0.4<x<1.0$ is accompanied by a strong-to-weak coupling crossover and that this shallow band remains involved in the superconducting pairing, although its contribution to the normal state fades away. Differences between hole- and electron-doped BaFe$_{2}$As$_{2}$ series are emphasized and discussed in terms of strong pair breaking by potential scatterers beyond the Born limit.    
\end{abstract}
\maketitle


%

\section{Introduction}

High-temperature superconductivity in the cuprates typically occurs at a crossover from a highly-correlated antiferromagnetic Mott insulating state to a weaker correlated Fermi liquid as a function of hole doping.~\cite{Lee06} In the 122 iron-pnictide superconductors, the Mott insulator is absent from the doping-temperature ($x$,T) phase diagram and the parent compound BaFe$_{2}$As$_{2} $ is metallic.~\cite{Johnston10,Stewart11,Georges13,Marinelli16} Thus, these systems were initially thought to be fairly weakly correlated materials. However, recent thermodynamic measurements on the fully K-substituted compound KFe$_{2}$As$_{2} $ have revealed that both the Sommerfeld coefficient $\gamma_{n}$ and the Pauli susceptibility are strongly enhanced with respect to their bare density-functional-theory (DFT) values.~\cite{Hardy13} In addition, quantum-oscillation (QO) experiments indicate that these correlations are even further enhanced in RbFe$_{2}$As$_{2} $ and CsFe$_{2}$As$_{2} $ for some selective bands.~\cite{Eilers15} Whereas correlations in cuprates originate from large values of the Hubbard $U$, theoretical works have stressed the particular relevance of Hund's coupling, $J_{H}$, and orbital selectivity in the iron pnictides to explain the origin of these correlation effects and the unconventional aspects of the metallic state.~\cite{Haule09,Hardy13,Georges13,deMedici14,Backes15} In particular, pioneering five-band Density-Functional Theory + Dynamical Mean Field Theory (DFT+DMFT) calculations~\cite{Haule09} predicted a coherence-incoherence crossover later found experimentally~\cite{Hardy13} and revealed that $J_{H}$ dramatically suppresses the coherence scale T$^{*}$ below which a metal with enhanced Pauli susceptibility is found, leaving an incoherent metal with local moments for T $>>$ T$^{*}$.  Thus, the Hund's coupling is responsible for the formation of the iron-local moment in these compounds consistent with the large fluctuating local moment on the Fe sites observed by x-ray emission spectroscopy.~\cite{Gretarsson11,Yin11}

In the cuprates, the superconducting energy gap $\Delta({\bf k})$ was proved experimentally to have $d_{x^{2}-y^{2}}$ symmetry~\cite{Tsuei00} and it is believed that the condensation of Cooper pairs is related to a spin-fluctuation exchange and not to a more conventional phonon mechanism. Indeed, the gap function changes sign between {\bf k} and {\bf k' = k+Q} on the single Fermi surface, where {\bf Q} = ($\pi,\pi$) is the momentum at which spin-fluctuation mediated pairing interaction $U({\bf k},{\bf k'})$ is peaked, in order to extract an attractive component from the screened Coulomb repulsion.~\cite{ChubukovAnnualReview} In iron pnictide superconductors, pairing is also probably due to spin fluctuations, but in this case {\bf Q} connects separated electron and hole sheets of the Fermi surface and $\Delta({\bf k})$ changes sign between these two sheets leading to a $s\pm$ state.~\cite{Hirschfeld11,ChubukovAnnualReview,Hirschfeld16} Since the structure of low-energy spin fluctuations evolves with doping, the same spin-fluctuation mechanism that gives rise to an $s\pm$ gap at moderate doping can theoretically give rise to a $d$-wave gap at stronger hole or electron doping via an intermediate $s+id$ state that breaks time-reversal symmetry.~\cite{Lee09,Stanev10,Khodas12,Platt12} In Ba$_{1-x}$K$_{x}$Fe$_{2}$As$_{2}$, the doping evolution of the superconducting-state symmetry is actually strongly disputed.~\cite{Ota14,Reid12b,Watanabe14} Indeed, near the optimal concentration ($x\approx0.35$), heat-capacity,~\cite{Popovich10} penetration-depth,~\cite{Cho16} and angle-resolved photoemission spectroscopy measurements (ARPES)~\cite{Ding08,Evtushinsky14,Shimojima11} unambiguously indicate nodeless $s$-wave energy gaps, while heat-transport data~\cite{Reid12} were interpreted as a $d_{x^{2}-y^{2}}$ state for $x=1.0$ in disagreement both thermodynamic and penetration-depth data.~\cite{Hardy14,Cho16} Thus, the symmetry of the superconducting state and its evolution with doping remains strongly debated in the Fe-based superconductors. 

Measurements of the specific heat play a crucial role in investigations of both the normal and superconducting properties. They provide direct information about the electronic density of states $N(0)$ that is not readily obtained from other techniques. Whereas most other measurements that give details about the energy gaps are sensitive to surface properties, {\it e.g.} penetration depth, ARPES and scanning tunneling spectroscopy (STM), heat capacity is a bulk property. As shown for {\it e.g.} MgB$_{2}$,~\cite{Fisher03,Fisher13} and cuprates~\cite{Wang01}, thermodynamic investigations (specific heat and thermal expansion) on doped samples of BaFe$_{2}$As$_{2}$ present a unique opportunity to study the effects of band filling and correlations as well as the superconducting-state symmetry and both the inter- and intraband couplings and scattering in multiband superconductors.

In this Article, we study both experimentally and theoretically the evolution of the normal- and superconducting-state thermodynamics of BaFe$_{2}$As$_{2}$ single crystals (i) with increasing hole content by K substitution and (ii) with isovalent substitution of K by Rb and Cs in KFe$_{2}$As$_{2}$. We find a huge enhancement of the Sommerfeld coefficient $\gamma_{n}$, proving that strong correlations strengthen with decreasing band filling and isovalent substitution. In addition, we find strong signatures of the coherence-incoherence crossover. These results can be understood, within an overall good agreement, by our Density-Functional-Theory + Slave Spin (DFT + SS) calculations, proving that iron pnictides are effectively Hund metals and that $J_{H}$ and orbital selectivity are the key ingredients for tuning correlations. As anticipated in Refs ~\onlinecite{Ishida10,deMedici14,Misawa12,Werner12}, hole doped pnictides are probably in the zone of influence of a Mott insulator that would be realized for half-filled conduction bands, {\it i.e.} for a doping of 1hole/Fe. This large quasiparticle mass enhancement is accompanied by a strong-to-weak coupling crossover of the superconducting-state for $x>0.40$, related to the disappearance of the electron sheets (shallow-band effect). No evidence for nodes in the energy gap are observed for any doping level, implying that there is no symmetry change of the Cooper-pair wavefunction in the overdoped region. Instead, a smooth decrease of all the energy gaps occurs beyond the optimal concentration, which simply correlates with the suppression of T$_{c}$. Finally, we emphasize the differences between hole- and electron-doped BaFe$_{2}$As$_{2}$ in terms of pair breaking which explains the apparent Bud'ko-Ni-Canfield (BNC)~\cite{Budko09,Budko13} scaling of the heat-capacity jump in electron-doped materials.\\

The Article is organized in the following way. In Sec.~\ref{methods} the experimental (crystal growth and thermodynamics) and theoretical (DFT + SS) methods are explained. In Sec.~\ref{results}, we present our raw specific-heat and thermal-expansion results. Details of the subtraction of the large lattice background from the specific-heat data are provided. The normal-state properties of Ba$_{1-x}$K$_{x}$Fe$_{2}$As$_{2}$ and Rb- and CsFe$_{2}$As$_{2}$ are discussed together with the theoretical calculations in Sec.~\ref{normal}. The doping evolution of the superconducting state of Ba$_{1-x}$K$_{x}$Fe$_{2}$As$_{2}$ is presented in Sec.~\ref{super} together with a comparison to electron-doped systems.
   
\section{Methods}\label{methods}

\subsection{Crystal growth and characterization}
High-quality single crystals of Ba$_{1-x}$K$_{x}$Fe$_{2}$As$_{2}$ were grown by a self-flux technique, using either FeAs or KAs fluxes, in alumina crucibles sealed in iron cylinders using very slow cooling rates of 0.2 - 0.4 $^{\circ}$ C/hour. All the crystals were annealed  {\it in-situ} by further slow cooling to room temperature. AFe$_{2}$As$_{2}$ single crystals (with A = Rb, Cs) were obtained under similar conditions using an As-rich flux. Samples with a typical mass of 1 to 5 mg were chosen for thermodynamic measurements and their composition was determined by refinement of four-circle single-crystal x-ray diffraction data of a small piece of each crystal. The high quality and the good homogeneity is attested by the sharp thermodynamic transitions, as illustrated in Figs~\ref{Fig1} and ~\ref{Fig4}, and by the recent observation of quantum oscillations (QO) in the magnetostriction of AFe$_{2}$As$_{2}$ (A = K, Rb, Cs) crystals.~\cite{Zocco13,EilersPHD,Eilers15} Moreover, our measurements yield bulk $T_{c}$ values of 2.5 K and 2.25 K for A = Rb, Cs, respectively which are 20 \% larger than reported in Refs~\onlinecite{Wang13} and~\onlinecite{Zhang15} .

\subsection{Thermodynamic measurements}
Thermal-expansion measurements were performed using a home-built capacitance dilatometer with a typical relative resolution $\Delta L/L \approx 10^{-8}-10^{-10}$.~\cite{Meingast90} Except for CsFe$_{2}$As$_{2}$, performing reproducible $c$-axis measurements proved to be quite difficult due to the large aspect ratios of the crystals, and we therefore present here mainly the in-plane measurements. Specific heat was measured using either the thermal relaxation~\cite{Bachmann72} or the dual-slope method~\cite{MarcenatPHD,Riegel86} in a Physical Property Measurement System from Quantum Design.

\subsection{Theoretical calculations}\label{SecTheory}
Calculations were carried out within the Density Functional Theory + Slave Spin technique (DFT + SS) for obtaining quasiparticle band structures renormalized by local dynamical electronic correlations. The ab-initio DFT bandstructures are calculated using the Generalized Gradient Approximation (GGA) for the exchange-correlation potential according to the Perdew-Burke-Ernzerhof recipe as implemented in Quantum Espresso~\cite{Giannozzi09} using the experimental lattice parameters and atomic positions. Wannier90~\cite{Mostofi08} is used to extract a local basis of Wannier functions for the five conduction bands of predominant Fe $3d$ character, allowing a tight-binding parametrization for these bands. In this basis the standard Kanamori Hamiltonian for Coulomb and Hund's coupling local interactions is used, 
\begin{eqnarray}\label{H_int}
 H_{int}&=& U\sum_{i,m}
 n^d_{im\uparrow} n^d_{im\downarrow}+U' \sum_{i,m>m',\sigma} n^d_{im\sigma} n^d_{im'\bar\sigma}\\
 &+&(U' -J_{H})\sum_{i,m>m',\sigma } n^d_{im\sigma} n^d_{im'\sigma},
\nonumber
\end{eqnarray}
where $n^d_{im\sigma}$ is the electron occupation number at site $i$ and for orbital $m$ and spin $\sigma$. $U$ and $U'=U-2J_{H}$ are the intra- and  inter-orbital Coulomb repulsions and $J_{H}$ is the Hund's coupling, treated here in the Ising (density-density only) form. The model is solved within the Slave-Spin mean-field approximation.~\cite{deMedici05,Hassan10} Further details on the method can be found in the Supplementary Material of Ref.~\onlinecite{deMedici14}. The value of the interaction parameters are set in the case of BaFe$_2$As$_2$, {\it i.e.} we use $U$ = 2.7 eV (a typical ab-initio calculated value), and $J_{H}/U$ = 0.25 (a realistic value in the semi-quantitative Slave-Spin mean-field with density-density only Hund's exchange interaction). These values are kept constant for all the calculations shown in this Article (albeit the interaction is believed to be somewhat stronger for K-, Rb- and CsFe$_2$As$_2$), in order to highlight the dependence of the correlation strength on doping and on the in-plane bandwidth for the isovalent compounds. The Sommerfeld coefficient is calculated from the total quasiparticle density of states at the Fermi level, $N(0)$, for the interacting system using,
\begin{equation}
\gamma_n=\pi^2k_B^2N(0)/3.
\end{equation}

\section{Results}\label{results}
\subsection{Heat-capacity and thermal-expansion measurements}
\begin{figure*}[t]
\begin{center}
\includegraphics[width=\textwidth]{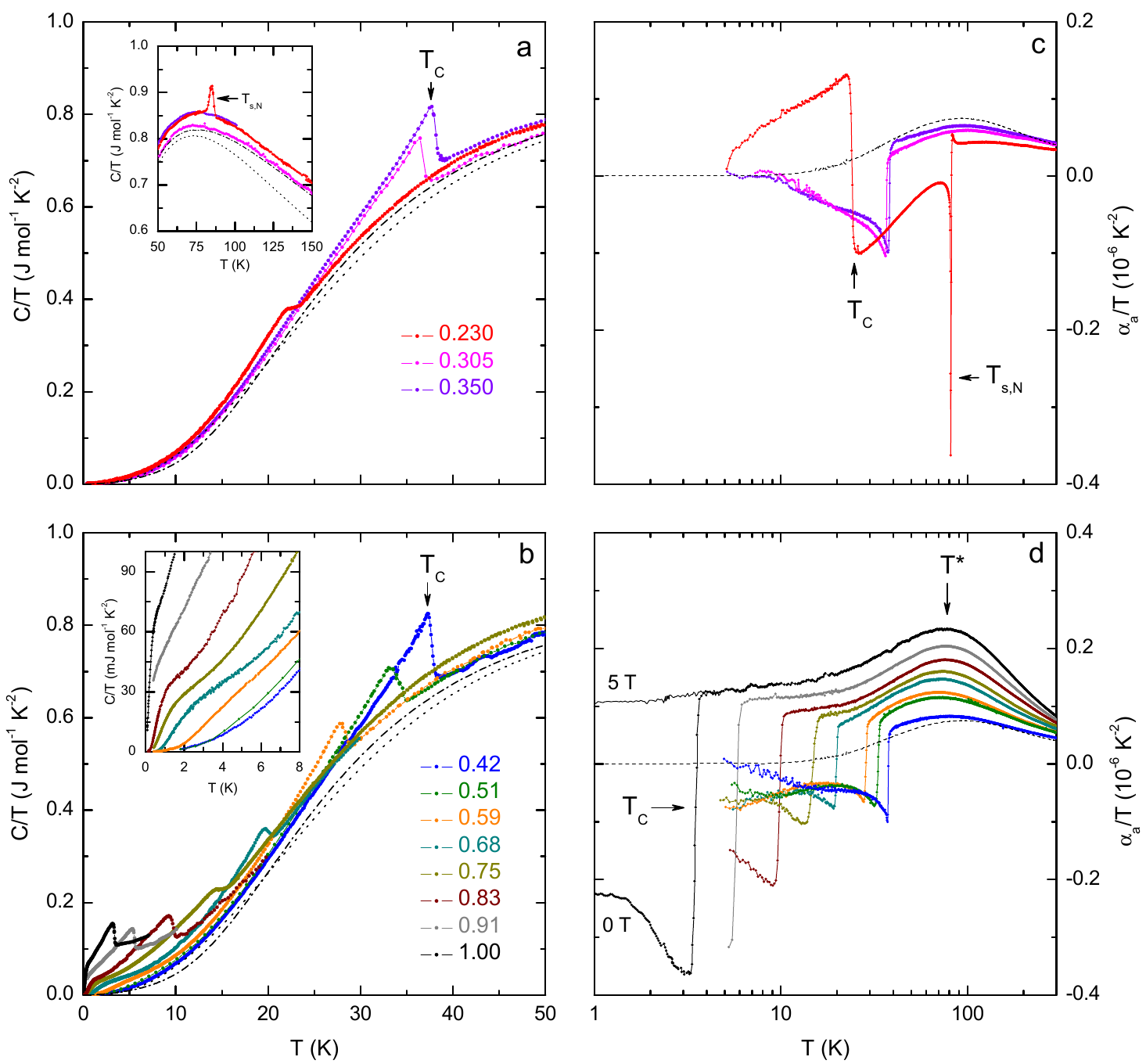}
\caption{\label{Fig1} (Color online) (a)-(b) Temperature dependence of the heat capacity of under- and overdoped Ba$_{1-x}$K$_{x}$Fe$_{2}$As$_{2}$ single crystals, respectively. The dotted and dash-dotted lines represent the lattice contributions of KFe$_{2}$As$_{2}$ and Ba(Fe$_{0.85}$Co$_{0.15}$)$_{2}$As$_{2}$, respectively, derived from Refs.~\onlinecite{Hardy10a} and~\onlinecite{Hardy13}. The insets show a magnification of the high- and low-temperature regions, respectively. (c)-(d) Temperature dependence of the in-plane thermal expansion of under- and overdoped Ba$_{1-x}$K$_{x}$Fe$_{2}$As$_{2}$ single crystals, respectively. The dashed line is the thermal expansion of Ba(Fe$_{0.67}$Co$_{0.33}$)$_{2}$As$_{2}$ taken from Ref.~\onlinecite{Meingast12}. The low-temperature thermal-expansion data of KFe$_{2}$As$_{2}$ (T $<$ 4 K) in 0 and 5 T were taken from Refs~\onlinecite{Burger13} and \onlinecite{Zocco13}.}
\end{center}
\end{figure*}

Figures~\ref{Fig1}a and~\ref{Fig1}b show the heat capacity and Figs~\ref{Fig1}c and~\ref{Fig1}d the in-plane thermal expansion of under- and overdoped Ba$_{1-x}$K$_{x}$Fe$_{2}$As$_{2}$ single crystals, respectively. Prominent anomalies in $C/T$ and $\alpha_{a}/T$ are clearly observed at both the SDW-structural ($T_{s,N}$) and SC transitions in both measurements. Superconducting transitions are quite narrow, with typically $\Delta T_{c}/T_{c}\leq 0.05$, reflecting the high homogeneity of these samples (except for the $x=0.23$ and $x=0.75$ compositions where $\Delta T_{c}/T_{c}$ is larger $\approx 0.12$ and 0.2, respectively). We note that the heat-capacity anomaly at $T_{c}$ reported in this work are substantially larger (by at least a factor of 1.2) and narrower than previously reported on polycrystalline samples ~\cite{Storey13,Budko13,Budko15}, reflecting the high-quality of our single crystals.

In Fig.~\ref{Fig1}, we have plotted both the specific heat and the thermal expansion divided by temperature since we focus on the electronic contributions in this paper. For a Fermi liquid, the electronic entropy $S_{e}(T)$ is linear in temperature resulting in constant electronic $C_{e}/T=\gamma_n$ and $\alpha_{e}/T$ contributions at low temperatures. The latter can be shown, via a Maxwell relation, to equal the pressure dependence of $\gamma_n$, so that,
\begin{equation}\label{defalpha}
\alpha_{e,i}/T=-\frac{1}{T}\left(\frac{\partial S_{e}}{\partial p_{i}}\right)=-\left(\frac{\partial \gamma_n}{\partial p_{i}}\right),
\end{equation}
where $i=\{a,c\}$.
As shown in Figs~\ref{Fig1}c and~\ref{Fig1}d, comparison with the expansivity of Ba(Fe$_{0.67}$Co$_{0.33}$)$_{2}$As$_{2}$,~\cite{Meingast12} which has the same crystal structure and a negligible electronic term, shows that the lattice contribution to $\alpha_{a}/T$ is quite small in Ba$_{1-x}$K$_{x}$Fe$_{2}$As$_{2}$ in comparison to the electronic term and becomes almost negligible for $x>0.42$. Thus, the low-temperature thermal-expansion signal is largely dominated by the electronic contribution in Ba$_{1-x}$K$_{x}$Fe$_{2}$As$_{2}$, particularly in overdoped samples.
\begin{figure*}[t]
\begin{center}
\includegraphics[width=\textwidth]{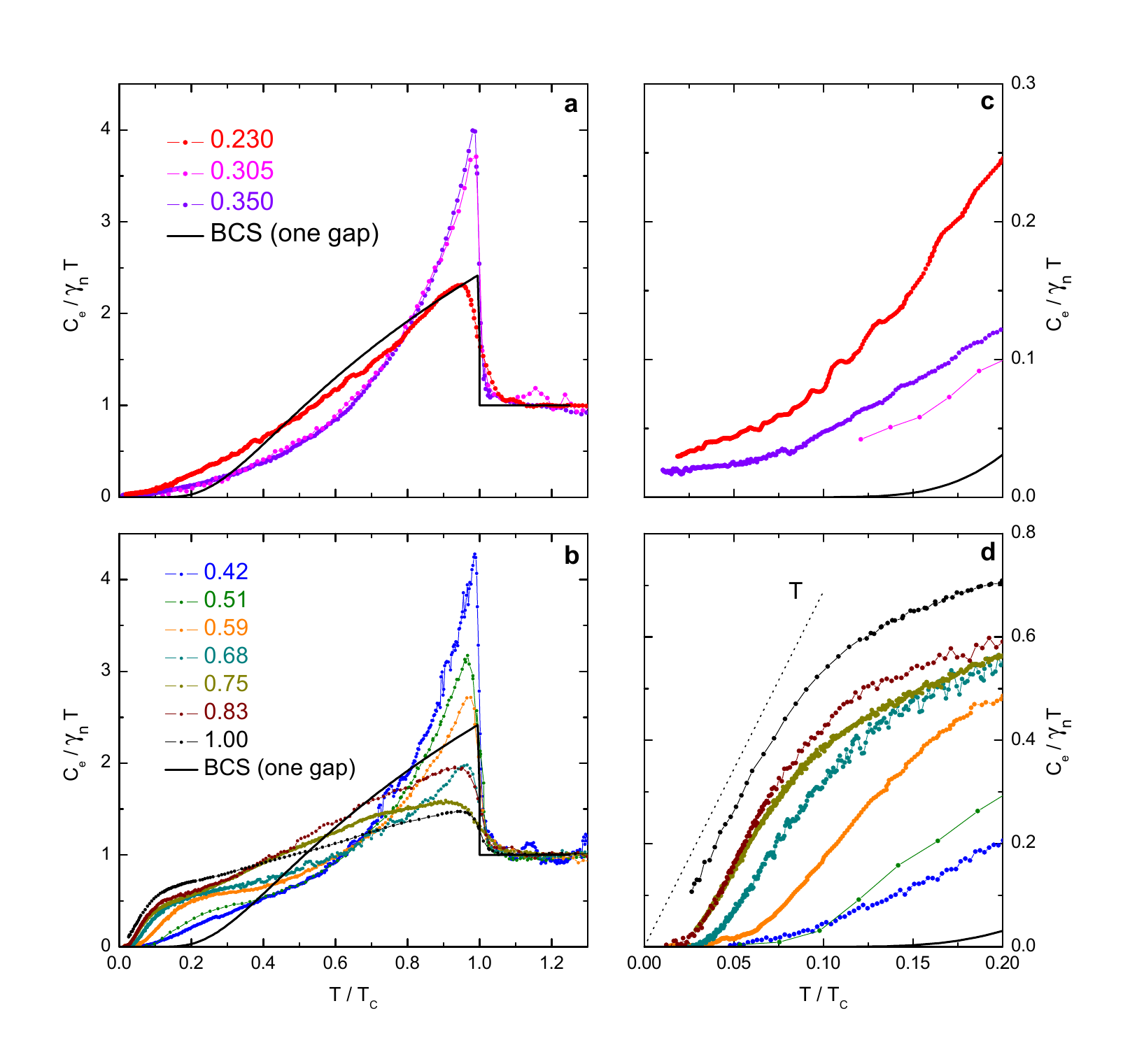}
\caption{\label{Fig2} (Color online) (a)-(d) Temperature dependence of the normalized electronic heat capacity, C$_{e}$/$\gamma_{n}$T, of under- and overdoped Ba$_{1-x}$K$_{x}$Fe$_{2}$As$_{2}$ single crystals, respectively. The solid curves represent the one-band weak-coupling BCS heat capacity for a $s$-wave superconductor. (c)-(d) Magnification of the low temperature region for under- and overdoped single crystals, respectively. The dotted line indicates a linear behavior expected for a nodal superconductor.} 
\end{center}
\end{figure*}
Conversely, the heat capacity is dominated by the lattice contribution $C_{L}(T)$, especially for compositions close to the optimal concentration ($x=0.35$, $T_{c}=38.4$ K), which complicates the thermodynamic analysis of the electronic properties of these compounds. Hereafter, we describe a reliable method to subtract the phonon background based on a modified version of the empirical Neumann-Kopp rule.~\cite{Gopal} 
\subsection{Extraction of the electronic specific heat}
We approximate the lattice specific heat C$_{L}(x,T)$ of Ba$_{1-x}$K$_{x}$Fe$_{2}$As$_{2}$ as the weighted sum of the individual lattice contributions of its 'constituents', BaFe$_{2}$As$_{2}$ and KFe$_{2}$As$_{2}$,~\cite{Qiu01} 
\begin{eqnarray}\label{Eq1}
C_{L}(x,T)&\approx(1-x)\cdot C_{L}(x=0,T)\nonumber\\
&+x\cdot C_{L}(x=1,T),
\end{eqnarray}
where $C_{L}(x=0,T)$ and $C_{L}(x=1,T)$ are the lattice specific heats of  BaFe$_{2}$As$_{2}$ and KFe$_{2}$As$_{2}$, respectively. Since BaFe$_{2}$As$_{2}$ is magnetic we use instead the lattice contribution of the non-superconducting Ba(Fe$_{0.85}$Co$_{0.15}$)$_{2}$As$_{2}$ derived in Ref.~\onlinecite{Hardy10a}, which was found to represent a reliable phonon background for Ba(Fe$_{1-x}$Co$_{x}$)$_{2}$As$_{2}$ in the range $0\leq x \leq 0.2$.~\cite{Hardy10b} On the other hand, $C_{L}(x=1,T)$ is inferred from the data in H = 5.5 T after subtraction of its electronic term, the latter being linear to about 50 K in KFe$_{2}$As$_{2}$.~\cite{Hardy13} Due to experimental uncertainties and in order to obtain an entropy-conserving electronic heat capacity $C_{e}(x,T)$, we have introduced a small correction factor $f_{s}\approx 0.98-1.02$ so that $C_{e}(x,T)=C(x,T)-f_{s}\cdot C_{L}(x,T)$ is obtained for each K content $x$ using Eq.~\ref{Eq1} for $C_{L}(x,T)$. The small deviation of $f_{s}$ from unity demonstrates that the above procedure represents a very good method to extract the electronic signal. We note that we have successfully applied the same method for Ba$_{1-x}$Na$_{x}$Fe$_{2}$As$_{2}$ in Ref.~\onlinecite{Wang16}. Our method is expected to be more reliable than using solely the lattice contribution of antiferromagnetic BaFe$_{2}$As$_{2}$ as a background, as done in Ref.~\onlinecite{Storey13}. In a previous report on polycrystalline samples with far broader superconducting transitions,~\cite{Kant10} an attempt was made to extract the electronic signal by fitting the normal state to the sum of Debye and Einstein contributions. However, this procedure leads to results that are inconsistent with both the actual bandstructure and the results of ARPES measurements.~\cite{Ding08,Evtushinsky14,Shimojima11}  

\begin{figure*}[t]
\begin{center}
\includegraphics[width=0.78\textwidth]{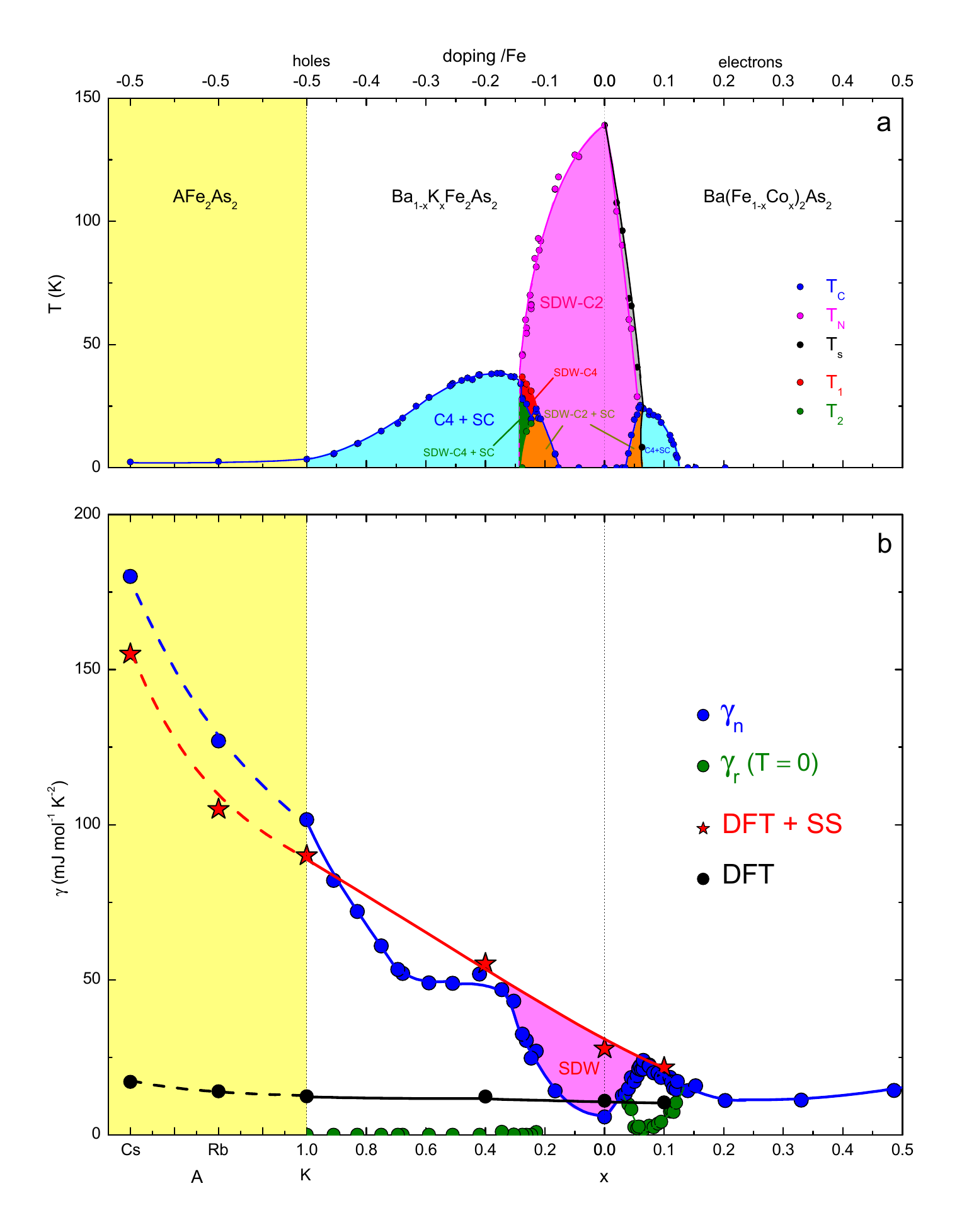}
\caption{\label{Fig3} (Color online)  (a) Phase diagram of Ba$_{1-x}$K$_{x}$Fe$_{2}$As$_{2}$ and Ba(Fe$_{1-x}$Co$_{x}$)$_{2}$As$_{2}$ derived from our specific-heat and thermal-expansion measurements.~\cite{Hardy10b,Bohmer15} $T_{c}$ of AFe$_{2}$As$_{2}$ (A = Rb, Cs) is also shown. $T_{1}$ and $T_{2}$ indicate the magnetic phase transitions SDW-C2 $\rightarrow$ SDW-C4 and SDW-C4 $\rightarrow$ SDW-C2 from Ref.~\onlinecite{Bohmer15}. (b) Sommerfeld coefficient $\gamma_{n}$ (blue symbols) and the residual density of states $\gamma_{r}$ at T = 0 K (green symbols) for Ba$_{1-x}$K$_{x}$Fe$_{2}$As$_{2}$, Ba(Fe$_{1-x}$Co$_{x}$)$_{2}$As$_{2}$ and AFe$_{2}$As$_{2}$ (A = Rb, Cs). The red stars represents DFT+SS calculations for the tetragonal paramagnetic phase (see text) and the pink area indicate the loss of density of states due to the reconstruction of the Fermi surface in the SDW phase. Lines are guide to the eyes.} 
\end{center}
\end{figure*}

In Fig.~\ref{Fig2}, we compare the resulting electronic contribution $C_{e}/\gamma_{n}T$, for various compositions ($0.23\leq x \leq1$), plotted as a function of $t \equiv T/T_{c}$. The inferred parameters $T_{s,N}$, $T_{c}$ and $\gamma_{n}$ are reported in Fig.~\ref{Fig3}, while the specific-heat jump ($\Delta C/\gamma_{n}T_{c}$) and the zero-temperature thermodynamic critical field ($H_{c}(0)$) are shown in Fig.~\ref{Fig7}.

\section{Discussion}
\subsection{The normal state of Ba$_{1-x}$K$_{x}$Fe$_{2}$As$_{2}$,  RbFe$_{2}$As$_{2}$ and CsFe$_{2}$As$_{2}$}\label{normal}
\subsubsection{Strong correlations and coherence-incoherence crossover}
Here we focus on the evolution of the Sommerfeld coefficient with hole and electron doping, as shown in Fig.~\ref{Fig3}b. On the Co-doped side, in the range $0.2<x<0.5$, $\gamma_{n}$ $\approx$ 12 mJ mol$^{-1}$K$^{-2}$ is minimum and the Fermi surface only consists of electron pockets.~\cite{Liu11} For smaller $x$, $\gamma_{n}$ rises significantly by a factor of two at the optimal concentration $x=0.06$. This increase is due to two concurring effects: on the one hand DFT calculations show that hole bands shift above the Fermi level, enhancing the bare density of states, and, on the other hand, the mass enhancement $\gamma_{n}/\gamma_{DFT}$ increases slightly from 1.7 at $x=0.2$ to 2.4 at optimal doping.\cite{Meingast12} We note that the mass enhancement observed near the optimal concentration is far smaller than that of BaFe$_{2}$(As$_{1-x}$P$_{x})_{2}$, where a value of $\gamma_{n}/\gamma_{DFT}\approx 10$ was obtained near optimal doping and interpreted as a sign of quantum criticality.~\cite{Walmsley13} For $x<0.06$, $\gamma_{n}$ decreases down to $\approx$ 6 mJ mol$^{-1}$K$^{-2}$ because of the Fermi-surface reconstruction induced by the SDW. 

On the hole-doped side, in the range $0<x<0.4$, $\gamma_{n}$ rises because the SDW is progressively suppressed by K substitution. By further increasing the hole concentration, $\gamma_{n}$ remains nearly constant up to $x\approx0.70$, with $\gamma_{n}\approx 50$ mJ mol$^{-1}$K$^{-2}$, corresponding to a mass enhancement of about 4 - 5. For $x>0.70$, $\gamma_{n}$ strongly increases and reaches 100 mJ mol$^{-1}$K$^{-2}$ for KFe$_{2}$As$_{2}$, a value comparable to moderate heavy-fermion systems.~\cite{Hardy13} 

Thus, omitting the SDW regions, we find that $\gamma_{n}$ grows continuously from $\gamma_{n}/\gamma_{DFT}$ $\approx$ 2 in the electron-doped side to about 7 - 9 in KFe$_{2}$As$_{2}$  demonstrating clearly that quasiparticles become heavier due to the strong correlations which develop with decreasing band filling. We note that our results  are in rough agreement with those of Ref.~\onlinecite{Storey13} obtained on polycrystals by differential calorimetry.

As shown in Figs~\ref{Fig1}c and~\ref{Fig1}d, this increase of the correlations with reducing band filling is accompanied by the emergence of a maximum of $\alpha_{a}/T$ around $T^{*}\approx75$ K for $x>0.42$, which is related to a coherence-incoherence crossover~\cite{Haule09} between a low-temperature heavy Landau Fermi liquid with a constant $\alpha_{a}/T$ and a high-temperature regime with a strongly reduced electronic thermal expansion. This identification is supported by the recent observation of a constant Pauli susceptibility for $T<<T^{*}$ and a Curie-Weiss-like behavior for $T>>T^{*}$ for compositions $x>0.47$.~\cite{Hardy13,Liu14} These observations are very reminiscent of the heavy-fermion behavior observed in $4f$ and $5f$ metals, where conduction electrons screen the local moments via the Kondo interaction leading to coherent heavy quasiparticles at low temperature. However, this dichotomy between localized and itinerant electrons is less evident in the iron pnictides where only $d$ electrons, forming multiple energy bands, are involved. In contrast to heavy fermion systems, we find here that $T^{*}$ does not scale with $\frac{1}{\gamma_{n}}$. Thus, neither the itinerant nor the local-moment approaches gives a satisfactory description of the normal state.~\cite{Gorkov13,YouBook} As discussed hereafter, this dichotomy is related to the coexistence of weakly (or light) and strongly correlated (or heavy) conduction electrons in Ba$_{1-x}$K$_{x}$Fe$_{2}$As$_{2}$ and this differentiation increases with both hole doping and isovalent substitution.~\cite{deMedici14,MediciBook} 

\subsubsection{Orbital selectivity and proximity to a Mott insulator}
\paragraph{Hole doping}
The observed coherence-incoherence crossover was predicted very early by DFT+DMFT calculations, and it was found that the coherence temperature $T^{*}$ is a strong function of the Hund's coupling constant $J_{H}$.~\cite{Haule09}  Unlike in the cuprates, it is $J_{H}$, rather than the Hubbard $U$, that determines the strength of the correlations in pnictides together with their multiorbital electronic structure.~\cite{MediciBook} Furthermore, it was demonstrated that, although the $d$ electrons are effectively itinerant, they do simultaneously contribute, due to this strong on-site Hund's interactions, to the large fluctuating local moment on the Fe sites observed by x-ray emission spectroscopy.~\cite{Gretarsson11,Yin11}

\begin{table*}[t] \caption{Schematic Fermi surface common to Rb-, CsFe$_{2}$As$_{2}$ and Ba$_{1-x}$K$_{x}$Fe$_{2}$As$_{2}$ ($x\geq0.4$). The table gives the band-resolved effective masses ($m^{*}$ in units of the bare electron mass) and the number of carriers ($n$ per Fe atom) for K-, Rb- and CsFe$_{2}$As$_{2}$ derived from quantum oscillations and our specific-heat measurements (warping of the different sheets is neglected). In Rb- and CsFe$_{2}$As$_{2}$, the $\beta$ orbit was not observed and (i) its mass is estimated using the measured Sommerfeld coefficient and (ii) its hole count is obtained assuming the same total Fermi volume as in KFe$_{2}$As$_{2}$. The dominant orbital character of each band is also given.}\label{tabQO}
\centering
\begin{tabular*}{\linewidth}{cp{0.10\linewidth}p{0.12\linewidth}p{0.12\linewidth}|p{0.12\linewidth}|p{0.12\linewidth}|p{0.12\linewidth}|p{0.09\linewidth}|p{0.09\linewidth}|p{0.09\linewidth}|p{0.09\linewidth}}
\multirow{7}{*}{\includegraphics[width=5.2cm]{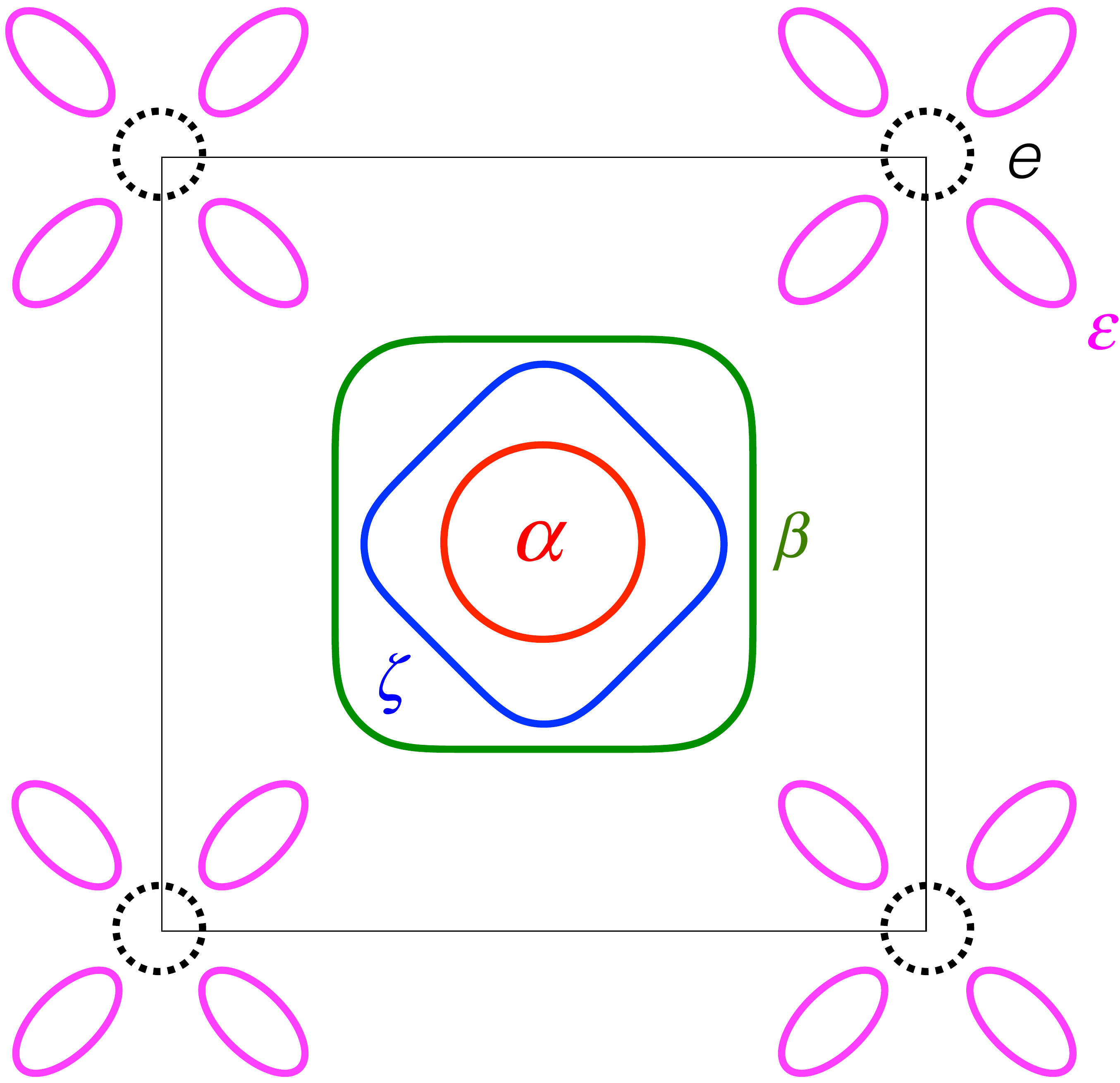}}&\multirow{12}{*}{}
\\
&&\multirow{12}{*}{}\\
\cline{2-11}
&\multicolumn {1}{c|}{\multirow{2}{*}{Band}}&\multicolumn {1}{c}{\multirow{2}{*}{Character}}&\multicolumn {2}{|c|}{$x=0.4$}&\multicolumn {2}{c|}{$x=1$}&\multicolumn {2}{c|}{RbFe$_{2}$As$_{2}$}&\multicolumn {2}{c}{CsFe$_{2}$As$_{2}$}\\
\cline{4-11}
&\multicolumn {1}{c|}{}		 &						&\multicolumn {1}{|c|}{$n$}&\multicolumn {1}{c|}{$m^{*}$}&\multicolumn {1}{c|}{$n$}&\multicolumn {1}{c|}{$m^{*}$}&\multicolumn {1}{c|}{$n$}&\multicolumn {1}{c|}{$m^{*}$}&\multicolumn {1}{c|}{$n$}&\multicolumn {1}{c}{$m^{*}$}\\
\cline{2-11}
&\multicolumn {1}{c|}{$\epsilon$}&\multicolumn {1}{c}{$xz/yz$}&\multicolumn {1}{|c|}{0.034} &\multicolumn {1}{c|}{1.40}&\multicolumn {1}{c|}{0.043}&\multicolumn {1}{c|}{6.6}&\multicolumn {1}{c|}{0.053}&\multicolumn {1}{c|}{8.0}&\multicolumn {1}{c|}{0.062}&\multicolumn {1}{c}{12.0}\\
\cline{2-11}
&\multicolumn {1}{c|}{$\alpha$}	  &\multicolumn {1}{c}{$yz$}    &\multicolumn {1}{|c|}{0.03} &\multicolumn {1}{c|}{4.80}&\multicolumn {1}{c|}{0.084}&\multicolumn {1}{c|}{6.3}& \multicolumn {1}{c|}{0.08}&\multicolumn {1}{c|}{6.0}&\multicolumn {1}{c|}{0.076}&\multicolumn {1}{c}{10.0}\\
\cline{2-11}
&\multicolumn {1}{c|}{$\zeta$}     &\multicolumn {1}{c}{$xz$}    &\multicolumn {1}{|c|}{0.03} &\multicolumn {1}{c|}{4.80}&\multicolumn {1}{c|}{0.132}&\multicolumn {1}{c|}{13.3}& \multicolumn {1}{c|}{0.114}&\multicolumn {1}{c|}{12.0}&\multicolumn {1}{c|}{0.121}&\multicolumn {1}{c}{19.0}\\
\cline{2-11}
&\multicolumn {1}{c|}{$\beta$}     &\multicolumn {1}{c}{$xy$}    &\multicolumn {1}{|c|}{0.11} &\multicolumn {1}{c|}{9.00}&\multicolumn {1}{c|}{0.257}&\multicolumn {1}{c|}{19.0}& \multicolumn {1}{c|}{0.268}&\multicolumn {1}{c|}{24.0}&\multicolumn {1}{c|}{0.254}&\multicolumn {1}{c}{41.0}\\
\cline{2-11}		
&\multicolumn {1}{c|}{$e$}	           &\multicolumn {1}{c}{$xz/yz$}	      &\multicolumn {1}{|c|}{0.005} &\multicolumn {1}{c|}{0.80}&\multicolumn {1}{c|}{0}&\multicolumn {1}{c|}{0}&	\multicolumn {1}{c|}{0}&\multicolumn {1}{c|}{0}&\multicolumn {1}{c|}{0}&\multicolumn {1}{c}{0}\\
\cline{2-11}
&\multicolumn {1}{c|}{Ref.}	  &\multicolumn {1}{c|}{\onlinecite{Yoshida14,Evtushinsky14}}&\multicolumn {2}{c|}{\onlinecite{Ding11,Evtushinsky09}}&\multicolumn {2}{c|}{\onlinecite{Terashima10,Terashima13,Zocco13,Yoshida14,Zocco14,Eilers15,EilersPHD}} 			&\multicolumn {2}{c|}{\onlinecite{Eilers15,EilersPHD}} &\multicolumn {2}{c}{\onlinecite{Eilers15,EilersPHD}} \\
\cline{2-11}
\multirow{12}{*}{}\\
\multirow{12}{*}{}\\
\end{tabular*}
\end{table*}

Interestingly, recent QO~\cite{Terashima10,Terashima13,Zocco13,Eilers15,EilersPHD} and ARPES experiments~\cite{Yoshida14}, summarized in Table~\ref{tabQO}, reveal that the mass enhancement is strongly orbital dependent in Ba$_{1-x}$K$_{x}$Fe$_{2}$As$_{2}$ for $x\geq0.4$. In particular, the outer hole sheet ($\beta$ band), with a dominant $xy$ character and the largest hole content, shows a significant mass enhancement $m^{*}/m_{e}$ $\approx$ 9 at $x=0.4$ that considerably increases to about 19 at $x=1.0$. Clearly, the degree of localization and, therefore the strength of the correlations, are strongly differentiated among the electrons forming the conduction bands, and this differentiation effectively increases with hole doping.\\

Recently, the strong increase of correlations with hole doping, the coherence-incoherence crossover and the orbital-selective mass enhancement reported here were all anticipated theoretically by several authors~\cite{deMedici14} and are interpreted by the proximity of hole-doped BaFe$_{2}$As$_{2}$ to a putative Mott insulating state that would be realized for half-filled conduction bands, {\it i.e.} for 1 hole/Fe. In this context, the end compound KFe$_{2}$As$_{2}$, with 0.5 hole/Fe, is thus located half way from it. De' Medici {\it et al.}~\cite{deMedici14,MediciBook} showed that our observations can be understood by the orbital decoupling mechanism induced by Hund's coupling, termed 'selective Mottness'. In this scenario, the Hund's coupling acts to decouple the different orbitals from each other and to increase the correlations selectively in each band implying that the main variable that tunes the correlations within each orbital is its doping with respect to individual half filling. Thus, it is the orbital population that determines the correlation strength in each orbital. In particular, it was found theoretically in Refs~\onlinecite{deMedici14,Hardy14} that the mass enhancement is the largest for the $xy$ band, which is the orbital closest to individual half filling (see Table~\ref{tabQO}).

In order to quantitatively check the validity of this model, we performed DFT + SS calculations for several compositions, as described in Section~\ref{SecTheory}, and compare them
to the experimental results in Fig.~\ref{Fig3}b. For $U$ = 2.7  eV and  $J_{H}/U$ = 0.25, the calculated $\gamma_{n}$ values are in good overall agreement with the experimental ones for both electron- and hole-doped compounds. These results indicate clearly that hole-doped pnictides are well described as hosting electrons in which the correlations are strongly orbital dependent, as it follows from the supposed influence of the nearby half-filled Mott insulator.
\begin{figure}[h]
\begin{center}
\includegraphics[width=9cm]{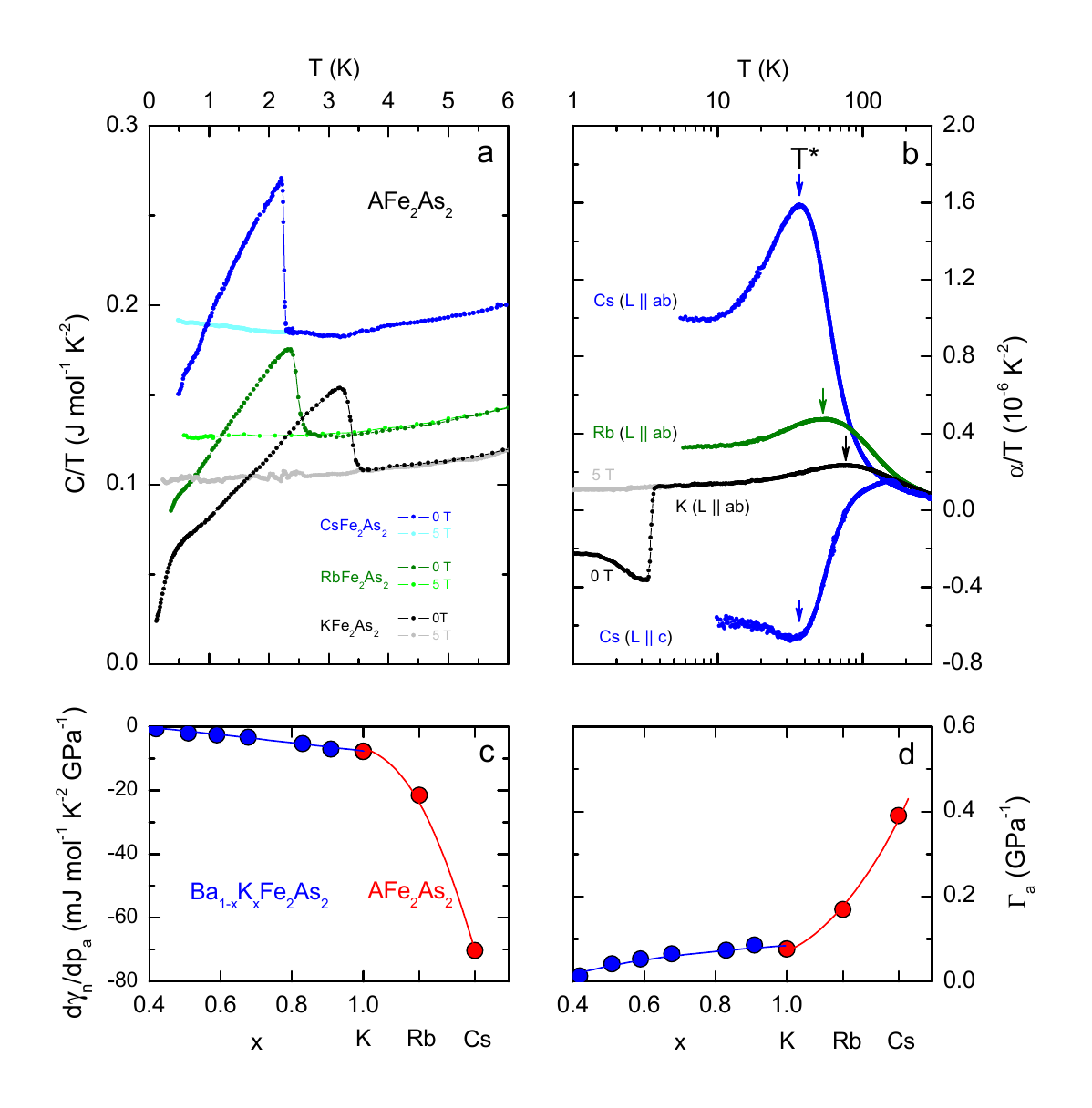}
\caption{\label{Fig4} (Color online) (a) Low-temperature specific heat of KFe$_{2}$As$_{2}$, RbFe$_{2}$As$_{2}$ and CsFe$_{2}$As$_{2}$ in 0 and 5 T. (b) Temperature dependence of the uniaxial thermal expansion of KFe$_{2}$As$_{2}$, RbFe$_{2}$As$_{2}$ and CsFe$_{2}$As$_{2}$. The arrows indicate the coherence temperature $T^{*}$. The low-temperature thermal-expansion data of KFe$_{2}$As$_{2}$ (T $<$ 4 K) in 0 and 5 T were taken from Refs~\onlinecite{Burger13} and~\onlinecite{Zocco13} . The dashed lines indicate the extrapolated low-temperature Fermi-liquid term. (c) - (d) Evolution of $d\gamma_{n}/dp_{a}$ and the Grüneisen parameter $\Gamma_{a}\propto-\frac{1}{\gamma_{n}}\frac{\gamma_{n}}{dp_{a}}$ for Ba$_{1-x}$K$_{x}$Fe$_{2}$As$_{2}$ ($0<x<1$) and AFe$_{2}$As$_{2}$ (A = K, Rb, Cs), respectively.} 
\end{center}
\end{figure}

\paragraph{Isovalent substitution}

To further test the relevance of this scenario, we performed additional heat-capacity and thermal-expansion measurements on the isovalent compounds RbFe$_{2}$As$_{2}$ and CsFe$_{2}$As$_{2}$, as shown in Figs~\ref{Fig4}a and b. We find that $\gamma_{n}$ considerably increases with the alkali radius (K $\rightarrow$ Rb $\rightarrow$ Cs), {\it i.e.} by stretching the unit cell, and reaches about 180 mJ mol$^{-1}$K$^{-2}$ for CsFe$_{2}$As$_{2}$, which is almost a factor 2 larger than in KFe$_{2}$As$_{2}$. Recent QO measurements~\cite{Eilers15,EilersPHD} confirm that the effective masses are strongly enhanced on all the Fermi-surface sheets, reaching {\it e.g.} $m^{*}/m_{e}$ $\approx$ 40 for the $xy$ band in CsFe$_{2}$As$_{2}$, while the individual band fillings remain unchanged within experimental accuracy (see Table~\ref{tabQO}). In Fig~\ref{Fig3}b, we compare our $\gamma$ values with those from our DFT + SS calculations obtained with the same values of $U$ and $J_{H}/U$ as before and using the experimental lattice parameters. Again, the agreement is fairly good, demonstrating that pnictides are Hund's metal and the relevance of selective Mott physics in these materials. We note that recent DFT + DMFT calculations also successfully identify the most strongly renormalized orbitals, although they underestimate the value of  $m^{*}/m_{e}$ in comparison to our more simple calculations.~\cite{Backes15}

Despite the heavier masses, Rb- and  CsFe$_{2}$As$_{2}$ are not closer to the Mott insulator than KFe$_{2}$As$_{2}$, because the mass enhancement alone is not a good measure of the charge localization, as shown recently in Ref.~\onlinecite{Fanfarillo15} and consistent with the results of Ref.~\onlinecite{deMedici11}. We note that our theoretical results are at odds with the DFT + SS calculations of Ref.~\onlinecite{Eilers15}. Indeed, they found the same value $\gamma_{n}\approx 100$ mJ mol$^{-1}$K$^{-2}$ for K-, Rb- and CsFe$_{2}$As$_{2}$ for the same parameters $U$ and $J_{H}/U$ used in our work. This deficiency was interpreted as a sign of antiferromagnetic quantum criticality, which was not taken into account in their model. In our case, the agreement with experiments is fairly good without having to invoke hypothetical critical fluctuations.

On the other hand, bond-length or volume changes offer another route towards stronger correlations and incoherence, as captured by our calculations and by recent DFT + DMFT calculations.~\cite{Backes15} Indeed, we find that the coherence-incoherence crossover becomes more prominent and that T$^{*}$, defined as the extremum in $\alpha/T$ in Fig.~\ref{Fig4}b, shrinks by a factor of about 2 between K- and CsFe$_{2}$As$_{2}$, while $\gamma_{n}$ is increased by the same amount. Thus, for the K, Rb, Cs series we recover that $T^{*}\propto\frac{1}{\gamma_{n}}$ which is a typical signature of heavy-fermion compounds. Similar to these materials, very strong uniaxial pressure dependences are observed. This is illustrated by the quite large values of both $d\gamma_{n}/dp_{i}$, $i=\{a,c\}$,  and the Grüneisen parameters $\Gamma_{i}\propto-\frac{1}{\gamma_{n}}\frac{d\gamma_{n}}{dp_{i}}$, inferred from the $T\rightarrow0$ limit of $\alpha_{i}/T$ (see Figs~\ref{Fig4}c,~\ref{Fig4}d and Table~\ref{tab1}).
\begin {table}[h]
\caption{Uniaxial pressure derivatives of $\gamma_{n}$ and Grüneisen parameters $\Gamma_{i}\propto-\frac{1}{\gamma_{n}}\left(\frac{d\gamma_{n}}{dp_{i}}\right)$, $i=\{a,c\}$, for K-, Rb- and CsFe$_{2}$As$_{2}$. Units are mJ mol$^{-1}$K$^{-2}$GPa$^{-1}$ and GPa$^{-1}$, respectively.} \label{tab1}
\centering
  \begin{tabular}{|c|c|c|c|c|}  
  \multicolumn {4}{c}{} \\
  \hline
  \hline
  $d\gamma_{n}/dp_{i}$&KFe$_{2}$As$_{2}$&RbFe$_{2}$As$_{2}$&CsFe$_{2}$As$_{2}$ \\
  \hline
  $a$ &-7&-22&-70 \\
  \hline
  $c$ & - & -&40 \\
  \hline
  \hline
  $volume$ &-&-&-100\\
  \hline
  \hline
  \hline
  $\Gamma_{i}$&KFe$_{2}$As$_{2}$&RbFe$_{2}$As$_{2}$&CsFe$_{2}$As$_{2}$ \\
  \hline
  $a$ &0.076&0.17&0.4 \\
  \hline
  $c$ & - & -&-0.22 \\
  \hline
  \hline
  $volume$ &-&-&0.58\\
  \hline
  \hline
 \end{tabular}
\end {table}
Moreover, our data for CsFe$_{2}$As$_{2}$ show that the effect of in-plane compression is larger than $c$-axis uniaxial pressure and opposite in sign. Thus, the other crucial parameter to tune the correlations, besides doping, is the Fe-Fe distance rather than the As height. Here, the larger mass enhancement is explained by a reduction of bandwidth related to a reduced hybridization of neighboring atomic orbitals, which is particularly drastic for bands having a dominant $xy$ character.~\cite{Backes15,Eilers15} On the other hand, these uniaxial pressure effects are significantly weaker in Ba$_{1-x}$K$_{x}$Fe$_{2}$As$_{2}$, as shown in Figs~\ref{Fig4}c and d.

\subsection{The superconducting state of Ba$_{1-x}$K$_{x}$Fe$_{2}$As$_{2}$ and Ba(Fe$_{1-x}$Co$_{x}$)$_{2}$As$_{2}$}\label{super}

\subsubsection{Multiband superconductivity and orbital selectivity}
\begin{figure*}[t]
\begin{center}
\includegraphics[width=0.8\textwidth]{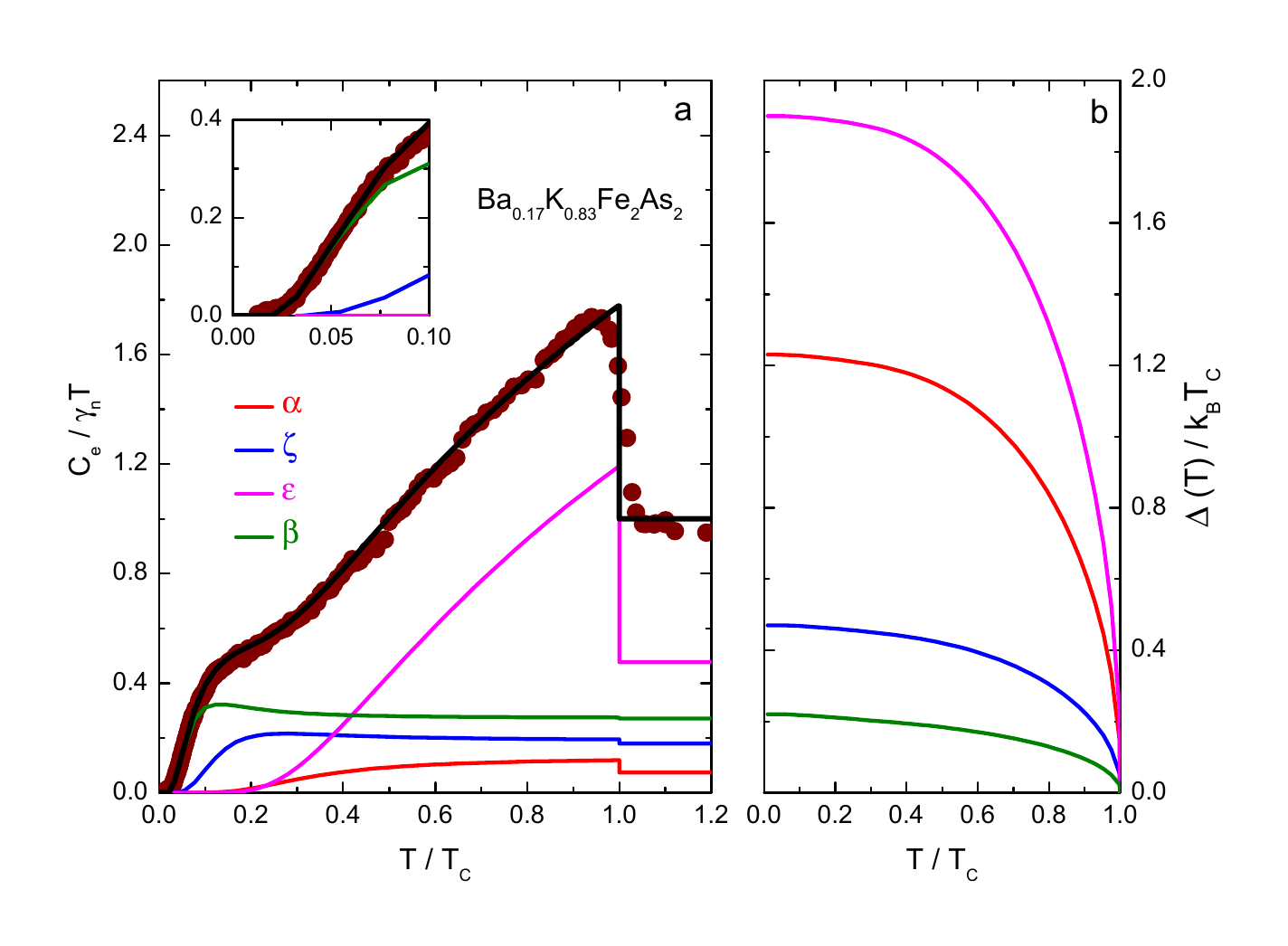}
\caption{\label{Fig5} (Color online) (a) Temperature dependence of the heat capacity of Ba$_{0.17}$K$_{0.83}$Fe$_{2}$As$_{2}$ derived in the 4-band isotropic BCS model (black line). Individual-band contributions are also shown. The inset shows a magnification of the low-temperature region, $T/T_{c}<0.1$. (b) Temperature dependence of the individual gaps obtained in this model.}  
\end{center}
\end{figure*}

\begin{figure}[t]
\begin{center}
\includegraphics[width=0.89\columnwidth]{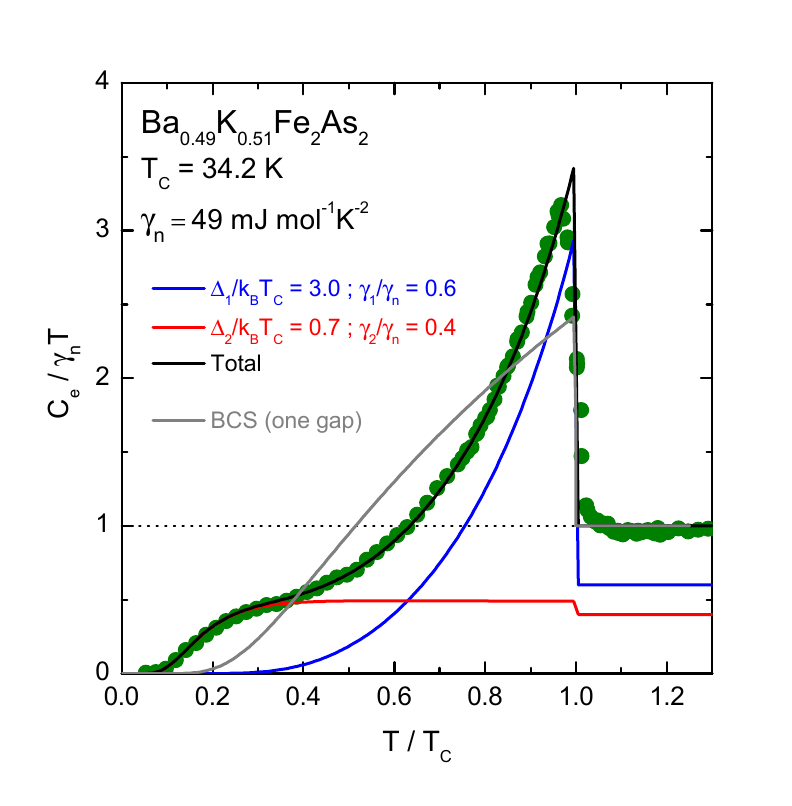}
\caption{\label{Fig6} (Color online) Electronic specific heat of Ba$_{0.49}$K$_{0.51}$Fe$_{2}$As$_{2}$ single crystal. The black curve represents a two-gap fit using the empirical two-band $\alpha$-model. The blue and red curves are the partial specific-heat contributions of the two bands.}
\end{center}
\end{figure}

\paragraph{Multiple energy gaps}
In Figs~\ref{Fig2}a and~\ref{Fig2}b , we clearly observe that $C_{e}/T$ significantly deviates from the single-band $s$-wave BCS behavior. At low temperature ($T/T_{c}<0.3$), $C_{e}/T$ is higher than the BCS curve, by orders of magnitude at the lowest temperature, for all K concentrations. This excess specific heat clearly indicates the presence of low-energy quasiparticle excitations, which are unambiguously related to the existence of small energy gaps, $\Delta_{S}(0)$, of amplitude significantly smaller than the single-band BCS value, $\Delta_{BCS}(0)=1.764k_{B}T_{c}$. As illustrated in Figs.~\ref{Fig2}c and ~\ref{Fig2}d, these curves exhibit a rapid increase of $C_{e}/T$ when the thermal energy $k_{B}T$ becomes of the order of $\Delta_{S}(T)$, and the smaller $\Delta_{S}(0)$ is the steeper is the increase of $C_{e}/T$. For $k_{B}T\gtrsim\Delta_{S}(T)$, the specific heat nearly reaches the normal-state value, although the system is still superconducting, as shown quantitatively in Fig.~\ref{Fig5}a (green and blue curves) and in Fig.~\ref{Fig6}a (red curve). At $T_{c}$, there is a small jump caused by the closing of $\Delta_{S}$ because the superconducting ground state is already almost empty and the amplitude of this residual jump is proportional to $\left(\frac{\Delta_{S}(0)}{k_{B}T_{c}}\right)^{2}$.

On the other hand, the positive curvature of $C_{e}/T$ for $T/T_{c}>0.5$ (see Figs~\ref{Fig2}a and~\ref{Fig2}b), where the BCS curve shows negative curvature, is an additional typical effect of multiband superconductors, that reflects the existence of at least one larger gap, $\Delta_{L}(0)>\Delta_{BCS}(0)$, as previously observed in MgB$_{2}$.~\cite{Fisher03,Fisher13} In this case, thermal excitation of the carriers across the gap occurs mainly in the vicinity of $T_{c}$, as illustrated quantitatively in Fig.~\ref{Fig5}a (magenta curve) and Fig.~\ref{Fig6} (blue curve).\\

\paragraph{Gap amplitudes}\label{gapamp}
A more quantitative description, providing quantitative values of the energy gaps, can be derived by analyzing the temperature dependence of $C_{e}(T)$ for $T<T_{c}$. This was successfully done for KFe$_{2}$As$_{2}$ in Ref.~\onlinecite{Hardy13} using a realistic weak-coupling 4-band BCS model exploiting the experimental band-resolved densities of states $N_{i}(0)$, $i=\{\alpha,\beta,\zeta,\epsilon\}$, inferred from QO and ARPES experiments.~\cite{Terashima10,Terashima13,Zocco13,Yoshida14,Zocco14,Eilers15,EilersPHD} Here, we find that this model applies equally well for the composition $x=0.83$ as demonstrated in Fig.~\ref{Fig5}. As input parameters, we used slightly different $N_{i}(0)$ values than those of KFe$_{2}$As$_{2}$, since they are not currently available for compositions $x<1$. The inferred gap amplitudes are compared to ARPES values in Table~\ref{tabgap}. We note that, unlike KFe$_{2}$As$_{2}$,~\cite{Hardy13} no $\cos(4\theta)$ modulation of the gaps is needed to reproduce precisely the measured $C_{e}(T)$ for $T/T_{c}<0.1$ (see inset of Fig.~\ref{Fig5}a). Our results are in good agreement with the ARPES results of Ref.~\onlinecite{Ota14} but are at odds with the unrealistically large values of Xu {\it et al.}~\cite{Xu13} ranging from 2.5 to 4.5 $k_{B}T_{c}$.
\begin{table}[t] \caption{Band densities of states $N_{i}(0)$ (given in units of the total density of states $N(0)$) used in the analysis of $C_{e}(T)$. $\Delta_{i}(0)$, given in units of $k_{B}T_{c}$, are the average gap amplitudes inferred from the 4-band BCS (for $x=1.0$ and $x=0.83$) and the empirical two-band $\alpha$ ($x=0.51$) models.}\label{tabgap}

\centering
\begin{tabular}{|c|c|c|c|c|c|c|}
\multicolumn {7}{l}{}\\
\hline
\hline
&	\multicolumn {2}{c|}{$x=1.0$}& \multicolumn {2}{c|}{$x=0.83$} & \multicolumn {2}{c|}{$x=0.51$}\\
\hline
\hline
&	\multicolumn {6}{c|}{C(T)} \\\hline
			&$N_{i}(0)$			&$\Delta_{i}(0)$			&$N_{i}(0)$		&$\Delta_{i}(0)$	&$N_{i}(0)$	&$\Delta_{i}(0)$\\
\hline
$\epsilon$ 	&0.36					&1.90					&0.48				&1.90			&\multirow{3}{*}{0.6}	&\multirow{3}{*}{3.0}\\
\cline{1-5}
$\alpha$ 		&0.10					&0.57					&0.07				&1.23			&				&\\
\cline{1-5}
$\zeta$ 		&0.23					&0.35 *					&0.18				&0.47			&				&\\
\hline
$\beta$ 		&0.31					&0.22					&0.27				&0.22			&0.4				&0.7\\
\hline
Ref.			&\multicolumn {2}{c|}{\onlinecite{Hardy14}}						&\multicolumn {2}{c|}{this work} 			&\multicolumn {2}{c|}{this work} \\			
\hline
\hline
&	\multicolumn {6}{c|}{ARPES} \\
\hline
 				&\multicolumn {2}{c|}{$\Delta_{i}(0)$}			&\multicolumn {2}{c|}{$\Delta_{i}(0)$}		&$N_{i}(0)$				&$\Delta_{i}(0)$	\\
\hline
$\epsilon$ 		&\multicolumn {2}{c|}{-}					&\multicolumn {2}{c|}{-}				&\multirow{3}{*}{0.63}		&\multirow{3}{*}{3.2}\\
\cline{1-5}
$\alpha$ 			&\multicolumn {2}{c|}{3.8}					&\multicolumn {2}{c|}{1.24}			&&\\	
\cline{1-5}
$\zeta$ 			&\multicolumn {2}{c|}{1.4*}					&\multicolumn {2}{c|}{0.66*}			&&\\
\hline
$\beta$ 			&\multicolumn {2}{c|}{0.5}					&\multicolumn {2}{c|}{0.66*}			&0.37		&<1.1\\
\hline
Ref.				&\multicolumn {2}{c|}{\onlinecite{Okazaki12}}		&\multicolumn {2}{c|}{\onlinecite{Ota14}} 	&\multicolumn {2}{c|}{\onlinecite{Evtushinsky09,Evtushinsky14,Ding08,Kordyuk12}} \\
\hline
\hline
\multicolumn {7}{l}{* with accidental nodes}\\
\end{tabular}
\end{table}
 
A similar self-consistent analysis is unfortunately not possible for other K compositions, because strong-coupling effects occur for $x<0.6$, as discussed hereafter in Section~\ref{strong}. However, reliable values of $\Delta_{i}(0)$ can still be obtained in the context of the multiband $\alpha$-model~\cite{Bouquet01} derived from Padamsee {\it et al.} strong-coupling model.~\cite{Padamsee73} In this approximation, the temperature dependence of the gaps is not obtained self-consistently but is rather taken to be the same as in the one-band weak-coupling BCS theory. The only adjustable parameters are the gap ratios $\Delta_{i}(0)/k_{B}T_{C}$ and the individual densities of states. As an example, Fig.~\ref{Fig6} shows that two isotropic gaps, of amplitude $\Delta_{1}(0)/k_{B}T_{c}\approx3.0$ and $\Delta_{2}(0)/k_{B}T_{c}\approx0.7$, can be inferred from the temperature dependence of $C_{e}/T$ for $x=0.51$. As shown in Table~\ref{tabgap}, these values are in excellent agreement with synchrotron ARPES measurements~\cite{Evtushinsky09,Evtushinsky14,Ding08,Kordyuk12}, which found the same gap amplitudes on the $\alpha$, $\xi$, $\epsilon$ and $e$ sheets and a significantly smaller one on the heaviest outer $\beta$ band. For completeness, we show in Fig.~\ref{Fig6a}a and~\ref{Fig6a}b, a 3-band analysis of the remaining compositions, and the inferred gaps and individual densities of states are illustrated in Fig.~\ref{Fig6a}c and~\ref{Fig6a}d, respectively. Evidence for important doping-induced changes in the superconducting-state properties are clearly observed. Coming from the underdoped side (see Fig.~\ref{Fig6a}c), the larger gap $\Delta_{3}(0)/k_{B}T_{c}$ initially increases and reaches its maximal value of $\approx$ 3.3 near $x=0.4$, and this can easily be understood by the suppression of the SDW state. However, for larger K content, $\Delta_{3}(0)/k_{B}T_{c}$ drops rapidly to 1.764 around $x\approx$ 0.7 and remains constant beyond this concentration. This feature is clearly not an artefact of the fitting procedure, since the specific-heat jump $\frac{\Delta C}{\gamma_{n}T_{c}}$, which is largely determined by the larger gap, exhibits also a singularity around $x$ $\approx$ $0.7$ (see Fig.~\ref{Fig7}c). In Sec.~\ref{strong} we argue that these features can be related to the disappearance of the electron band and represent a manifestation of the highly debated 'shallow-band effect'.~\cite{Bang14,Koshelev14,Hu15,Chen15,Leong15,Gorkov15} 

Similar to KFe$_{2}$As$_{2}$,~\cite{Hardy14} our analysis demonstrates that the larger gap largely determines the jump at $T_{c}$ while the $T\rightarrow0$ behavior is exclusively governed by several smaller gaps, $\Delta_{S}(0)$, for all compositions. Thus, our data show clearly two smooth trends with increasing K content above $x=0.42$: (i) a strong decrease of the jump height ($\Delta C/\gamma_{n}T_{c}$) and (ii) a steeper increase of the low-T $C_{e}/T$ with temperature, as illustrated in Figs~\ref{Fig2}d and~\ref{Fig6a}, for $T/T_{c}<0.2$. These features are both readily explained by the simultaneous decrease of the larger gap (from $\approx$ 3.3 to 1.9 $k_{B}T_{c}$) and several smaller gaps that accompany the suppression of $T_{c}$ with increasing $x$ beyond optimal doping.\\
\paragraph{Orbital selectivity and pairing}
Orbital selectivity is not only crucial for understanding the normal state but also for superconducting pairing. As pointed out experimentally by Evtushinsky {\it et al.}~\cite{Evtushinsky14} for $x=0.42$, and by Okazaki {\it et al.}~\cite{Okazaki12} for $x=1.0$, our analysis is also consistent with the fact that the smaller gap is always found on the heaviest outer hole $\beta$ band with $xy$ character. This indicates that this band is not actively involved in pairing. On the other hand, pairing is effectively quite strong for the more mobile $xz/yz$ electrons. 

This strong orbital sensitivity of pairing substantiates an intraorbital mechanism, because Hund's coupling strongly suppresses interorbital fluctuations.~\cite{deMedici14} These observations are compatible with a spin-fluctuations mediated mechanism characterized by dominant interband electron-pair scattering between parts of the Fermi-surface sheets having the same orbital $xz/yz$ character.~\cite{Kuroki08,Kuroki09}

\subsubsection{Absence of gap nodes}
\begin{figure*}[t]
\begin{center}
\includegraphics[width=0.82\textwidth]{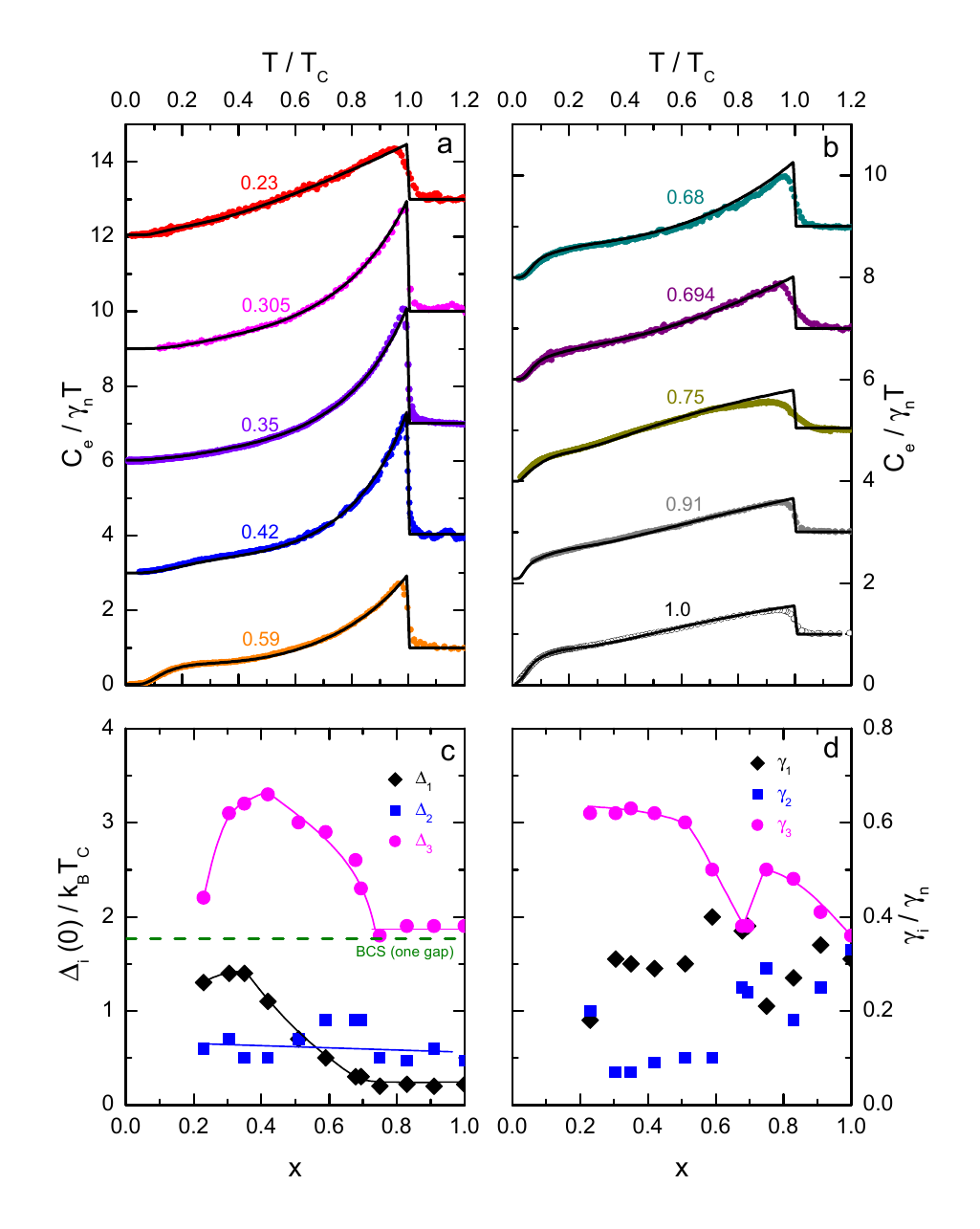}
\caption{\label{Fig6a} (Color online) (a)-(b) 3-band analysis (solid line) of the superconducting-state heat capacity of Ba$_{1-x}$K$_{x}$Fe$_{2}$As$_{2}$. For clarity, curves are vertically shifted by 3(2) from each other. (c)-(d) Inferred gap and individual densitiy-of-states values.}
\end{center}
\end{figure*}

As shown in Fig.~\ref{Fig2}d,~\ref{Fig6}a and~\ref{Fig6a}, $C_{e}$/T at low temperature is dominated by quite small energy gaps and vanishes exponentially to zero in the limit $T \rightarrow 0$ rather than linearly, as expected for line nodes in $\Delta(\bf{k})$.~\cite{Hardy14} Thus, our data exclude simultaneously the change of symmetry from $s-$ to $d$-wave between $0.35<x<1.0$ and the proposed change from $d$ to $s$ in KFe$_{2}$As$_{2}$ under pressure, as put forward in Refs.~\onlinecite{Reid12} and~\onlinecite{Tafti13}, respectively. We note that this scenario was also invalidated by recent penetration-depth measurements.~\cite{Cho16} Our data are also at odds with recent laser ARPES experiments, in which accidental nodes were found on several Fermi-surface sheets for $0.76<x<1.0$.~\cite{Xu13,Ota14} Furthermore, the absence of a sizable residual density of states, $\gamma_{r}$, in the $T\rightarrow 0$ limit, shows that our single crystals are insensitive to out-of-plane disorder induced by K substitution. This is in excellent agreement with the negligibly small residual $\kappa(0)/T$ reported in the recent heat-transport study of Hong {\it et al.}~\cite{Hong14} in the range $0.8<x<1.0$. It is however in striking contrast to Ba(Fe$_{1-x}$Co$_{x}$)$_{2}$As$_{2}$, for which sizable doping-dependent $\gamma_{r}$ and $\kappa_{0}/T$ values are found away from the optimal concentration $x=0.06$ (see Fig.~\ref{Fig3}b).~\cite{Hardy10b,Gofryk11,Mu10,Reid10} Clearly, in-plane disorder is much more detrimental to superconductivity, as confirmed by the fast suppression of $T_{c}$ in K(Fe$_{1-x}$Co$_{x}$)$_{2}$As$_{2}$ and the 10 K reduction of $T_{c}$ in the optimally-doped Ba$_{1-x}$K$_{x}$(Fe$_{0.93}$Co$_{0.07}$)$_{2}$As$_{2}$ ($x\approx0.45$) with respect to that of Ba$_{0.65}$K$_{0.35}$Fe$_{2}$As$_{2}$.~\cite{Tafti13,Zinth11} A quantitative description of pair breaking induced by impurity scattering is presented in Section~\ref{impurity}.

\subsubsection{Strong-to-weak coupling crossover and the disappearance of the electron pockets}\label{strong}

\begin{figure}[t]
\begin{center}
\includegraphics[width=0.90\columnwidth]{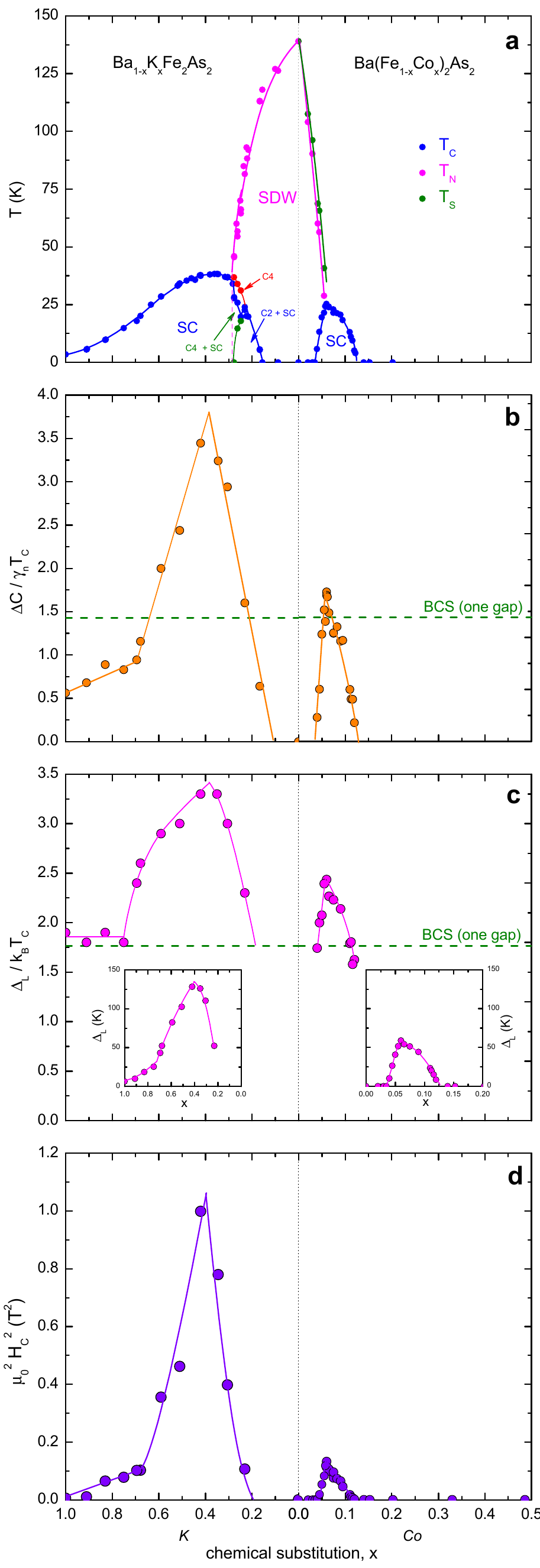}
\caption{\label{Fig7} (Color online) (a) Phase diagram of Ba$_{1-x}$K$_{x}$Fe$_{2}$As$_{2}$ and Ba(Fe$_{1-x}$Co$_{x}$)$_{2}$As$_{2}$ derived from our specific-heat and thermal-expansion measurements. (b) Evolution of the specific-heat jump at $T_{c}$. (c) Larger energy gaps derived from the temperature dependence of $C_{e}/T$. (d) Evolution of the zero-temperature thermodynamic critical field derived from our specific-heat measurements. Data for Ba(Fe$_{1-x}$Co$_{x}$)$_{2}$As$_{2}$ are taken from Refs~\onlinecite{Hardy10b,Meingast12}. The green line indicates the weak-coupling single-band BCS value.} 
\end{center}
\end{figure}

In the BCS weak-coupling limit, the value of the specific-heat discontinuity at $T_{c}$, $\frac{\Delta C}{\gamma_{n}T_{c}}$, in a two-band $s$-wave superconductor is always less than the single-band BCS value, $\left(\frac{\Delta C}{C_{n}}\right) _{BCS}$ = 1.426, and the energy gaps obey the supplemental condition, $\Delta_{S}(0)<1.764k_{B}T_{c}<\Delta_{L}(0)$.~\cite{Moskalenko59,Soda66,Geilikman67,Kresin73,Kresin90} As illustrated in Figs~\ref{Fig6a}c,~\ref{Fig7}b and~\ref{Fig7}c, these two conditions are simultaneously fullfilled only for $x>0.7$ in Ba$_{1-x}$K$_{x}$Fe$_{2}$As$_{2}$. For $0.2<x<0.7$, $\frac{\Delta C}{\gamma_{n}T_{c}}$ largely surpasses this threshold value reaching a maximum value of  $2.4\times\left(\frac{\Delta C}{C_{n}}\right) _{BCS}$ at the optimal concentration with large energy gaps $\Delta_{L} $ exceeding $3.0k_{B}T_{c}$ as shown in Figs.~\ref{Fig7}b and ~\ref{Fig7}c. In parallel, the Cooper-pairs condensation energy, $g_{n}(0)-g_{S}(0)=\mu_{0}H_{c}^{2}(0)/2$, inferred from our data (see Fig.~\ref{Fig7}d), is also strongly enhanced. We interpret this in terms of  a crossover from a weak- to a strong-coupling regime around optimal doping.~\cite{Carbotte90}
Coming from the overdoped side, these strong-coupling effects start to appear for $x\leq0.7$, {\it i.e.} approximately where a Lifshitz transition~\cite{Lifshitz60} is expected to mark the incipience of the electron pockets at the Fermi level. The effects of incipient or shallow bands on the superconducting properties are highly discussed both experimentally~\cite{He13,Miao15} and theoretically~\cite{Bang14,Koshelev14,Hu15,Chen15,Leong15,Gorkov15} in pnictide superconductors. As shown by Bang~\cite{Bang14} and Koshelev~\cite{Koshelev14}, this electron pocket, which progressively disappears as a function of K substitution, still plays a role in superconducting pairing even though the bottom of this band, E$_{g}$, is shifted above the Fermi energy. Thus, this shallow or empty band in the normal state can still display an energy gap below $T_{c}$ via pair hopping with the deeper hole bands, as long as $-\omega_{c}<E_{g}<\omega_{c}$, where $\omega_{c}$ is the high-energy cut-off of the pairing interaction. We show, hereafter, that this scenario can qualitatively explain our experimental data.

\begin{figure}[htb]
 \centering
\includegraphics[width=0.89\columnwidth]{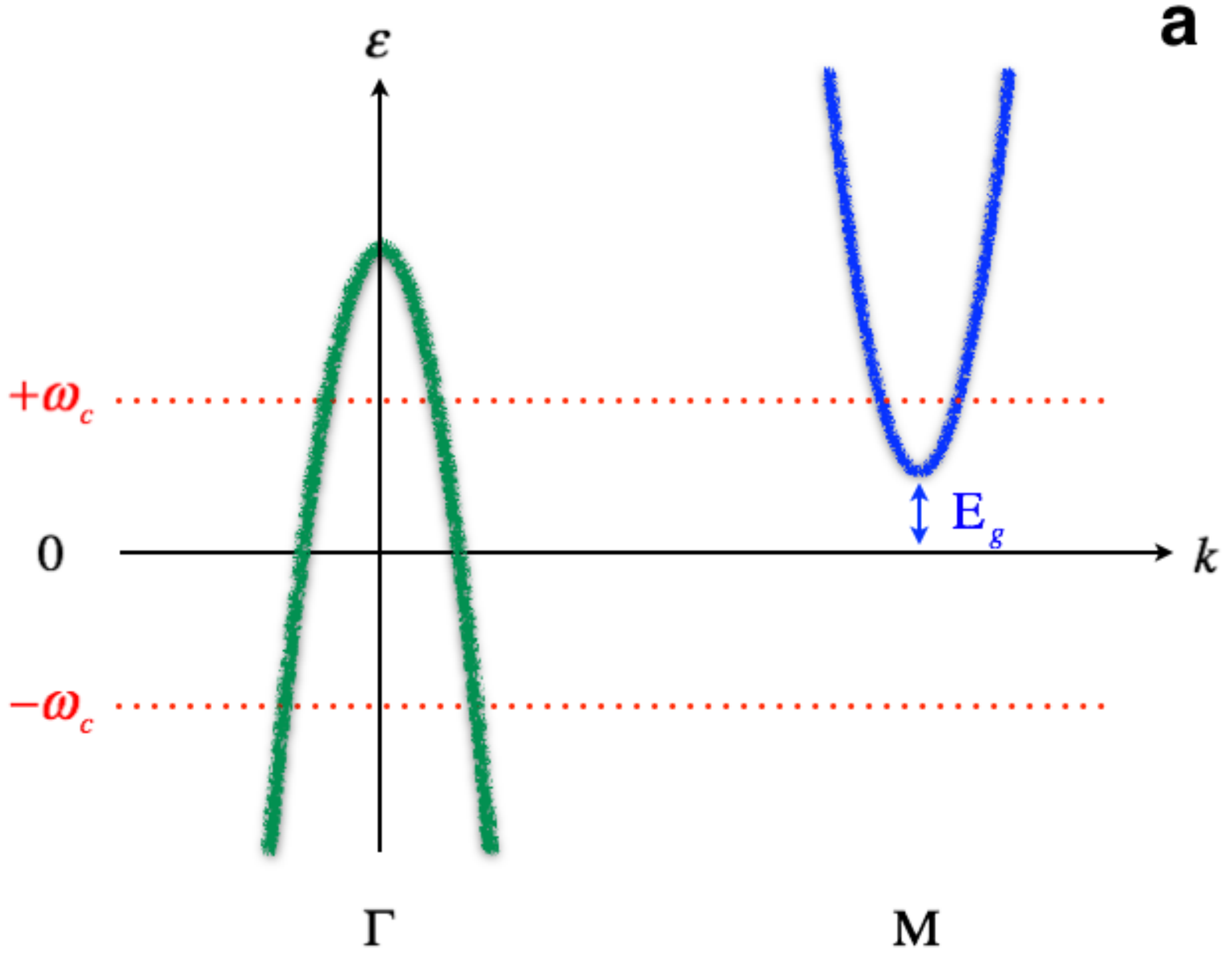}
\includegraphics[width=0.89\columnwidth]{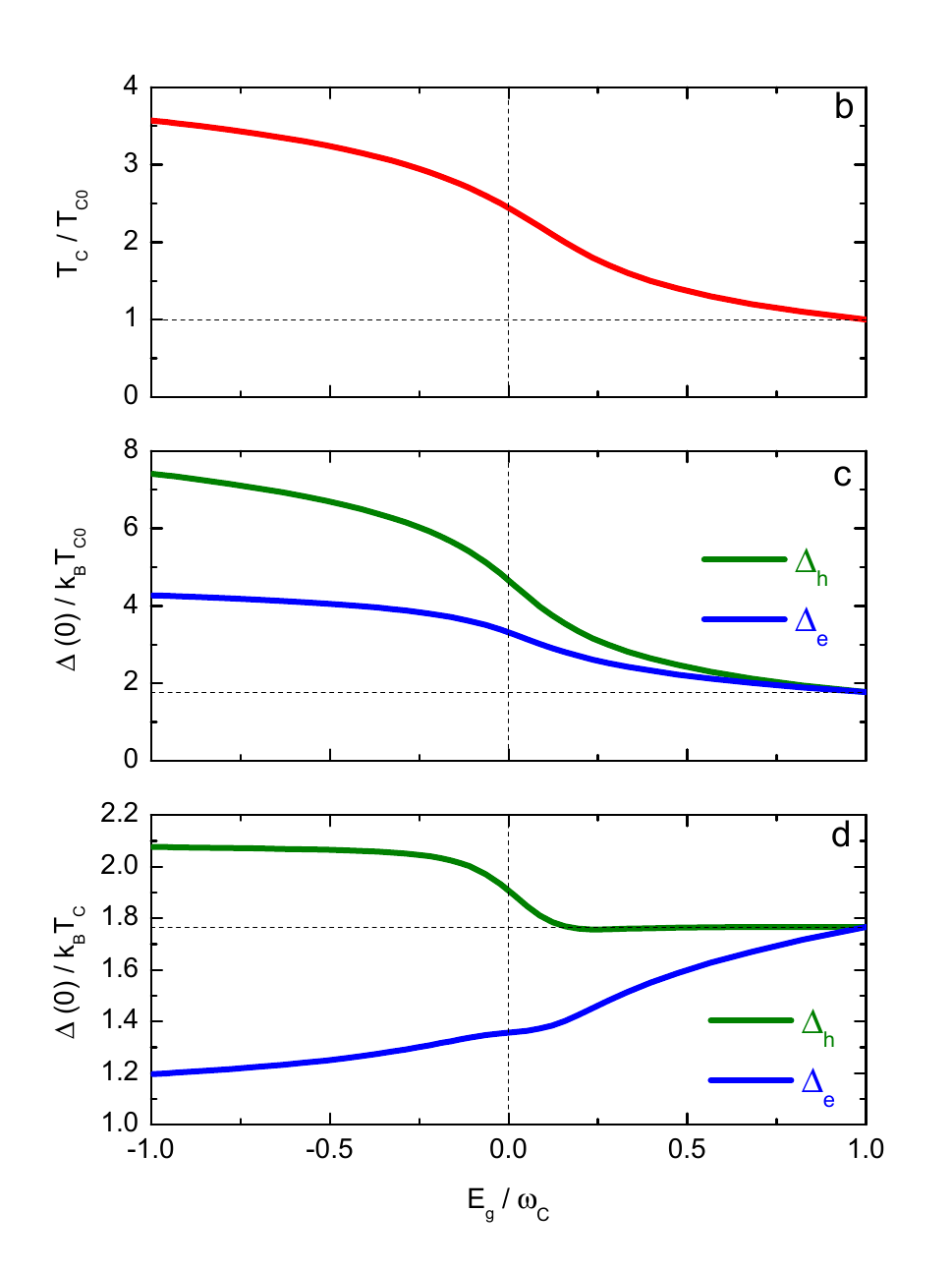}
\caption{\label{FigIncipient} (Color online) (a) Schematics of the two-band model. The hole band (in green) is a deep band. $E_{g}$ measures the distance of the bottom of the incipient electron band (in blue) from the Fermi energy. $\omega_{c}$ is the high-energy cut-off of the pairing interaction. (b)-(c) Evolution of $T_{c}$ and the zero-temperature energy gaps as a function of $E_{g}$. Both are normalized to $T_{c0}$, the transition temperature for the one-band case, {\it i.e.} for $E_{g}/\omega_{c}>1$. (d) Zero-temperature energy gaps normalized by $T_{c}$ as a function of $E_{g}$.}
\end{figure}

We consider a simplified 2D-model, similar to that of Refs ~\onlinecite{Bang14,Koshelev14,Chen15}, with one deep hole band and a shallow electron band within the weak-coupling BCS theory,  as illustrated in Fig.~\ref{FigIncipient}a. In this context, the gap equations are,
\begin{equation}\label{incipient}
  \left\{
      \begin{aligned}
          \Delta_{h}&=&\lambda_{hh}\,\Delta_{h}\,\chi_{h}&\,+\lambda_{he}\,\Delta_{e}\,\chi_{e}\\
          \Delta_{e}&=&\lambda_{ee}\,\Delta_{e}\,\chi_{e}&\,+\lambda_{eh}\,\Delta_{h}\,\chi_{h}\\   
             \end{aligned}
    \right.
\end{equation} 
with
\begin{equation}\label{incipient2}
\chi_{h}=\int_{-\omega_{c}}^{\omega_{c}}d\epsilon\frac{\tanh\left(\beta\frac{E_{h}}{2}\right)}{2E_{h}}\textrm{ and }\chi_{e}=\int_{E_{g}}^{\omega_{c}}d\epsilon\frac{\tanh\left(\beta\frac{E_{e}}{2}\right)}{2E_{e}}.
\end{equation}  
Here, $E_{h,e}=\sqrt{\epsilon^{2}+\Delta_{h,e}^{2}}$ and $\Delta_{h}$ and $\Delta_{e}$ are the energy gaps on the hole and electron sheets, respectively. $\lambda_{ij}=V_{ij}N_{j}$, with $i=\{h,e\}$, represent the dimensionless intraband ($i \neq j$) and interband ($i=j$) pairing strengths and $\beta=\frac{1}{k_{B}T}$. We set $\lambda_{ee}=0$ so that superconductivity on the electron band is induced only via pair hopping with the deeper hole band, and for simplicity, we choose $\lambda_{hh}=\lambda_{eh}=\lambda_{he}=0.3$.

In Figs~\ref{FigIncipient}b and~\ref{FigIncipient}c, we show the evolution of $T_{c}$ and the zero-temperature gaps, $\Delta_{e,h}(0)$, as a function of E$_{g}$, obtained by solving Eqs~\ref{incipient} in the limit $\Delta_{h,e}\rightarrow0$ and $T\rightarrow0$, respectively. For $E_{g}/\omega_{c}<-1$, the electron band is deep below the Fermi energy and the system behaves like a conventional two-band superconductor. By increasing $E_{g}$, $T_{c}$ and both gaps decrease smoothly in absolute value (see Figs~\ref{FigIncipient}b and c), and the conventional BCS one-band case is recovered for $E_{g}/\omega_{c}>1$. Our model reproduces qualitatively the suppression of $T_{c}$ observed experimentally for $x>0.4$ in Ba$_{1-x}$K$_{x}$Fe$_{2}$As$_{2}$, as illustrated in Fig.~\ref{Fig7}a. The inflexion point, found near $x\approx0.7$ in our data, can be identified as the locus of the Lifshitz transition ($E_{g}=0$). Moreover, our calculation show that the normalized hole gap, $\Delta_{h}(0)/T_{c}$, also smoothly decreases through the Lifshitz transition and saturates at a value of 1.764 for $E_{g}>0$ (see Fig.~\ref{FigIncipient}d). This explains qualitatively the evolution of the larger gap $\Delta_{L}(0)/T_{c}$ inferred from our data for $x>0.4$, as indicated in Fig.~\ref{Fig6a}c and Fig.~\ref{Fig7}c. In our model, superconductivity in the hole band is of intraband origin ({\it e.g.} phonon-like), while interband pairing is likely of electronic origin ({\it e.g.} spin fluctuations). Thus, coming from the overdoped side, the increase of $T_{c}$, $\frac{\Delta C}{\gamma_{n}T_{c}}$ and $\Delta_{L}(0)/T_{c}$ towards optimal doping can be understood as due to the bootstrap of electron-phonon superconductivity via spin fluctuations by coupling the incipient and the regular bands, as discussed in Ref.~\onlinecite{Chen15}.

In contrast, the above model is not appropriate for Ba(Fe$_{1-x}$Co$_{x}$)$_{2}$As$_{2}$. Here, $\frac{\Delta C}{\gamma_{n}T_{c}}$ slightly exceeds the BCS value only near $x=0.06$ in Ba(Fe$_{1-x}$Co$_{x}$)$_{2}$As$_{2}$ (see Fig.~\ref{Fig7}b) and drops rapidly away from optimal doping, a behavior that anticorrelates with $\gamma_{r}(x)$.~\cite{Hardy10b,Gofryk11,Mu10} This behavior cannot be explained by the existence of a parasitic second phase as argued in Ref.~\onlinecite{Kim12}, and $\gamma_{r}$ can only be accounted for by the existence of in-gap states ~\cite{Glatz10,Bang09} induced by the strong scattering of Cooper pairs by the Co dopant, {\it i.e.} beyond the Born limit, in the case of $s\pm$ symmetry.~\cite{Gofryk11} Indeed, scanning-tunneling spectroscopy (STS) measurements on Ba(Fe$_{1-x}$Co$_{x}$)$_{2}$As$_{2}$~\cite{Teague11} revealed a large excess zero-bias conductance, which is absent in Ba$_{0.58}$K$_{0.42}$Fe$_{2}$As$_{2}$.~\cite{Shan12}

\subsubsection{Relevance of the Bud'ko-Ni-Canfield (BNC) scaling ?}

\begin{figure}[h]
\begin{center}
\includegraphics[width=\columnwidth]{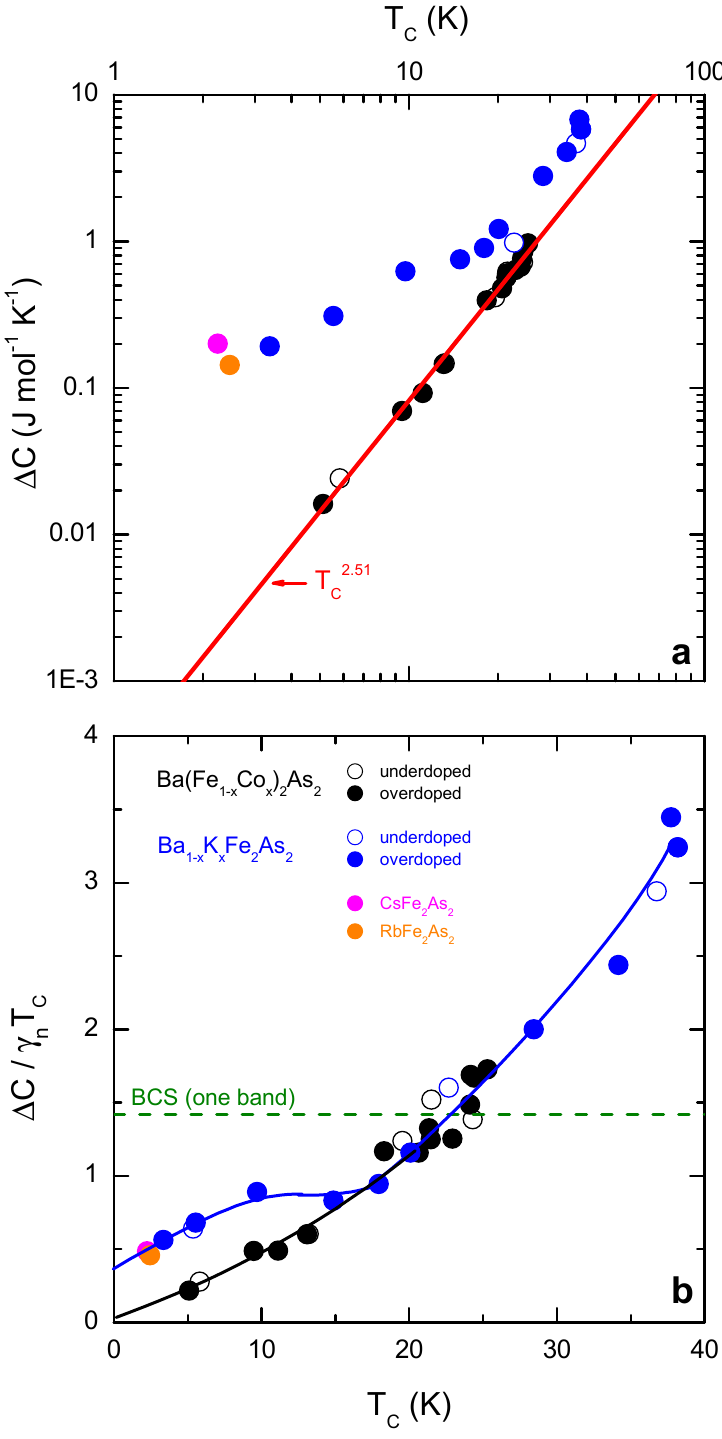}
\caption{\label{Fig8} (Color online) (a) $\Delta C/T_{c}$ as a function of $T_{c}$ in the log-log representation of Bud'ko-Ni-Canfield for under- (open symbols) and overdoped (closed symbols) single crystals of Ba(Fe$_{1-x}$Co$_{x}$)$_{2}$As$_{2}$ and Ba$_{1-x}$K$_{x}$Fe$_{2}$As$_{2}$. (b) $\Delta C/\gamma_{n}T_{c}$ as function of $T_{c}$. The dashed line indicates the weak-coupling single-band BCS value. Data for Ba(Fe$_{1-x}$Co$_{x}$)$_{2}$As$_{2}$ are taken from Refs~\onlinecite{Hardy10b,Meingast12}. Solid lines are guides to the eye.} 
\end{center}
\end{figure}

In Refs~\onlinecite{Budko09,Budko13,Budko14,Budko15}, it was reported that the specific-heat jump $\Delta C (T_{c})$ of single crystals of Ba(Fe$_{1-x}$TM$_{x}$)$_{2}$As$_{2}$ (TM = Co, Ni), polycrystals of Ba$_{1-x}$K$_{x}$Fe$_{2}$As$_{2}$ (for $x<0.7$) and Ba$_{1-x}$Na$_{x}$Fe$_{2}$As$_{2}$, follows a 'universal' trend $\Delta C (T_{c})\propto  T_{c}^{3}$. This scaling behavior was interpreted as arising from quantum criticality~\cite{Zaanen09} or from impurity-induced pair-breaking effects in the Born limit~\cite{Kogan09}. In Fig.~\ref{Fig8}a, we plot $\Delta C$ as a function of $T_{c}$ for all our single crystals of Ba$_{1-x}$K$_{x}$Fe$_{2}$As$_{2}$, Ba(Fe$_{1-x}$Co$_{x}$)$_{2}$As$_{2}$, RbFe$_{2}$As$_{2}$ and CsFe$_{2}$As$_{2}$ in the log-log representation of Bud'ko, Ni and Canfield (BNC). Our results differ from those of Refs~\onlinecite{Budko09,Budko13,Budko14,Budko15}, because only data for under- and overdoped Ba(Fe$_{1-x}$Co$_{x}$)$_{2}$As$_{2}$ appear to fall on a single 'scaling' curve. Also, we find an exponent $n=2.5$ and not 3.

A physically more relevant quantity to plot is $\frac{\Delta C}{\gamma_{n}T_{c}}$ as a function of $T_{c}$ on a linear scale (see Fig.~\ref{Fig8}b), because this takes the changes of density of states into account. $\Delta C=a\gamma_{n}T_{c}$ is a measure of both the normal-state electronic entropy $\gamma_{n}T_{c}$ and the strength of the superconducting coupling $a$, which can take any value in clean or dirty multiband superconductors. For Ba(Fe$_{1-x}$Co$_{x}$)$_{2}$As$_{2}$, we find that $\Delta C/\gamma_{n}T_{c}\rightarrow 0$ in the limit $T_{c}\rightarrow 0$ in line with the aforementioned existence of finite density of states $\gamma_{r}$ in the limit $T\rightarrow0$. This can only be explained by the destruction of Cooper pairs due to in-plane disorder induced by substitution of Fe by Co. Remarkably, both under- and overdoped values fall on the same curve indicating that $\Delta C/\gamma_{n}T_{c}$ is rather insensitive to the presence of the SDW state. On the other hand, $\Delta C/\gamma_{n}T_{c}$ tends to a finite value $\approx$ 0.35 in overdoped Ba$_{1-x}$K$_{x}$Fe$_{2}$As$_{2}$, RbFe$_{2}$As$_{2}$ and CsFe$_{2}$As$_{2}$. Here $\gamma_{r}$ is always zero, as expected for clean superconductors. This proves that the out-of-plane disorder does not alter substantially superconductivity in these systems. Below, we show quantitatively that pair-breaking effects are important in Ba(Fe$_{1-x}$Co$_{x}$)$_{2}$As$_{2}$ in accord with a $s\pm$ ground state. 

\subsubsection{Pair breaking and $s\pm$ superconductivity in Ba(Fe$_{1-x}$Co$_{x}$)$_{2}$As$_{2}$}\label{impurity}

Here, we will show that the dependence of $\frac{\Delta C}{\gamma_{n}T_{c}}$ on $T_{c}$ in Ba(Fe$_{1-x}$Co$_{x}$)$_{2}$As$_{2}$ can be quantitatively explained by the nonmagnetic scattering by Co atoms, beyond the Born limit, for a $s\pm$ state.

\begin{figure}[h]
\begin{center}
\includegraphics[width=\columnwidth]{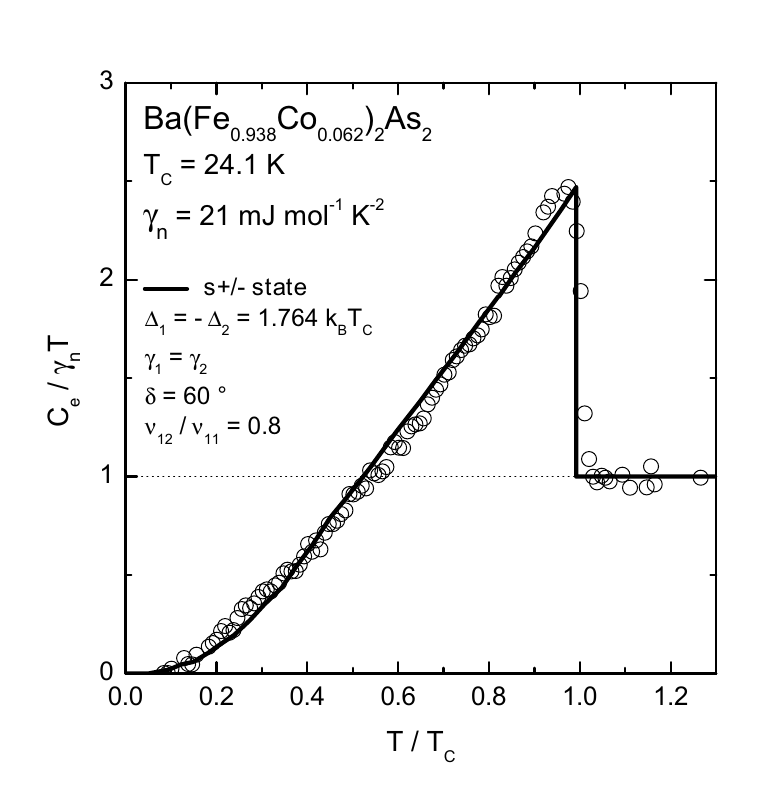}
\caption{\label{Fig11} Electronic specific heat of the optimally doped Ba(Fe$_{0.938}$Co$_{0.062}$)$_{2}$As$_{2}$ single crystal. The black curve is a calculation for an impure $s\pm$ superconductor (with equal densities of states on the hole and electron Fermi surfaces, $\gamma_{1}$ = $\gamma_{2}$) in the self-consistent $T$-matrix approximation (taken from Ref.~\onlinecite{Gofryk11}). The fitting parameters are the Friedel phase shift $\delta$ and the inter- to intraband scattering amplitudes ratio.} 
\end{center}
\end{figure}

We consider the simplified model of Gofryk {\it et al.}~\cite{Gofryk11} of a $s\pm$ ground state, which consists of only two bands with equal densities of states, $N_{1}(0)=N_{2}(0)=N(0)$, and energy gaps of opposite sign, $\Delta_{1}=-\Delta_{2}=\Delta$. As illustrated in Fig.~\ref{Fig11}, this model describes fairly well the temperature dependence of $C_{e}/T$ of our optimal Co concentration, $x=0.06$, with an intermediate scattering strength indicated by the Friedel phase shift $\delta$ = 60$^{\circ}$ and a significant interband scattering measured by $\nu_{12}/\nu_{11}=0.8$ (see Ref.~\onlinecite{Gofryk11} for more details on the calculations). 

In the following, we further simplify this model by neglecting intraband scattering, which is not pair-breaking in a $s\pm$ state. For this particular case, several authors ~\cite{Preosti94,Golubov97,Matsumoto09,Glatz10} have shown that the problem of interband scattering by potential scatterers in a $s\pm$ superconductor leads to the same equations for the Green's function and $T$-matrix as a conventional single-band $s$-wave superconductor with 'classical spins' as described by Yu, Shiba and Rusinov.~\cite{Yu65,Shiba68,Rusinov69,Shiba73,Balatsky06} This identification allows us to use the analytical formulas derived by Chaba and Singh Nagi~\cite{Chaba72} to fit our experimental data.

\begin{figure}[t]
\begin{center}
\includegraphics[width=\columnwidth]{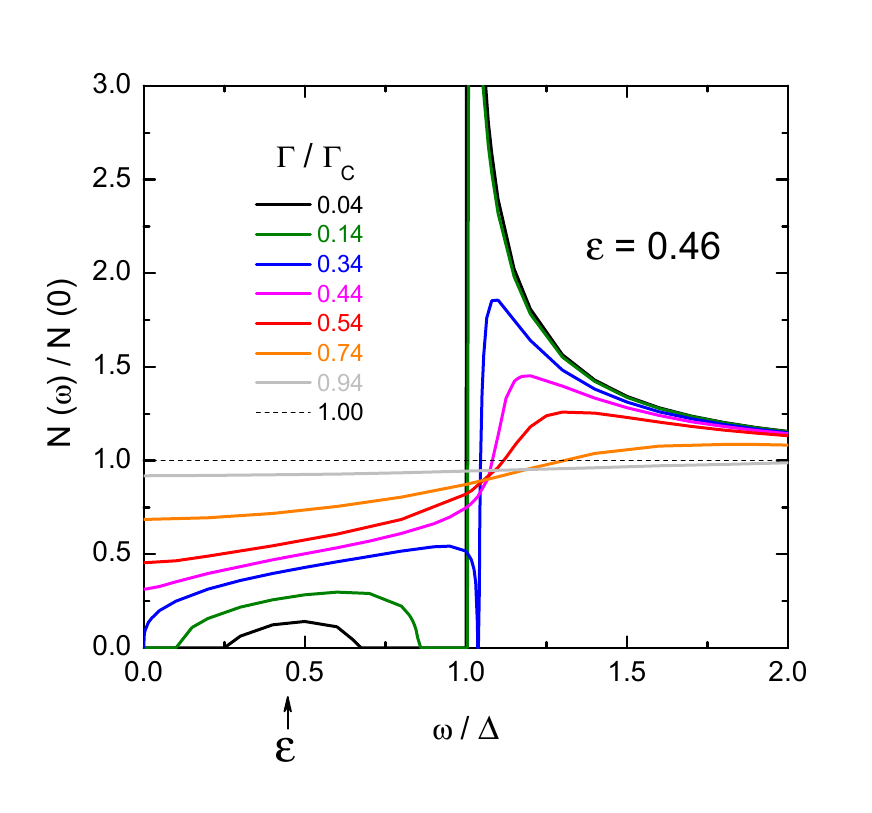}
\caption{\label{Fig9} (Color online) The density of states of quasiparticle excitations for a dirty $s\pm$ superconductor ($N_{1}(0)=N_{2}(0)=N(0)$, $\Delta_{1}=-\Delta_{2}=\Delta$) with nonmagnetic impurities as a function of the quasiparticle energy, for different values of the scattering rate $\Gamma/\Gamma_{c}$. All curves were calculated with $\epsilon$ = 0.46 (or equivalently $\delta$ = 60$^{\circ}$). Gaplessness occurs for $\Gamma/\Gamma_{c}>0.34$. Here $\Delta$ stands for $\Delta(\Gamma,T)$.}   
\end{center}
\end{figure}

In this context, the scattering rate $\Gamma$ is defined as 
\begin{equation}\label{scattering}
\Gamma=\frac{n_{i}}{2\pi N(0)}  \underbrace{\left(1-\epsilon^{2}\right)}_{\sin^{2}\delta},
\end{equation}
with $n_{i}$ the density of impurities and $N(0)$ the density of states (for one spin orientation ) which, in the present case, corresponds to the concentration of Co atoms, $x$, and the measured values of $\gamma_{n}$, respectively. Here, $\epsilon$ is the position of the bound state inside the energy gap induced by the scattering of the electrons by the Co atoms, as illustrated in Fig.~\ref{Fig9}. Thus, $\epsilon$ (or equivalently $\delta$) measures the strength of the impurity potential. The weak scattering limit (Born limit) corresponds to $\epsilon \rightarrow 1$ ($\delta\approx0$) and the bound state appears at the gap edge. In the opposite limit ({\it i.e.} unitary limit), $\epsilon \rightarrow 0$ ($\delta\approx\pi/2$) and the bound state appears near the Fermi level. As illustrated in Fig.~\ref{Fig9}, increasing the concentration of impurities ({\it $\Gamma$}) increases the number of bound states (which then form an impurity band), reduces the size of the gap leading first to a gapless state and ultimately to the complete suppression of superconductivity.

\begin{figure}[t]
\begin{center}
\includegraphics[width=0.90\columnwidth]{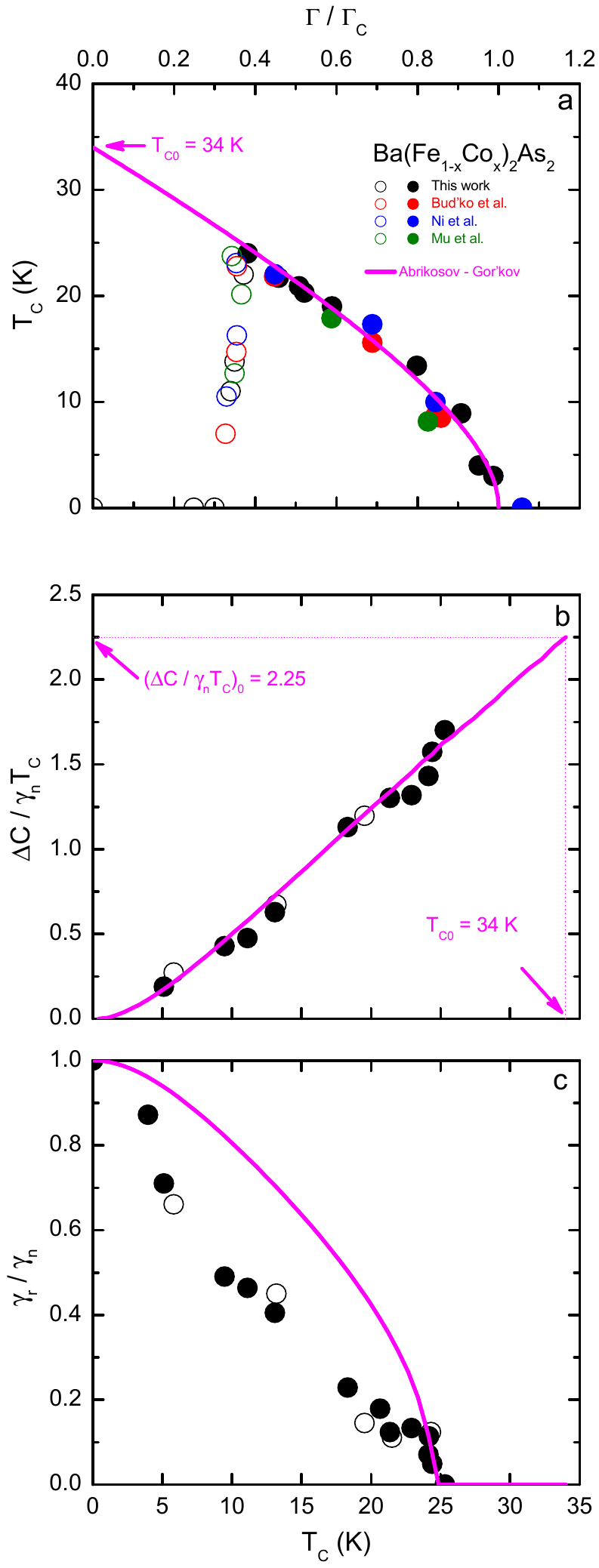}
\caption{\label{Fig10} (Color online) (a) Evolution of $T_{c}$ on the scattering rate $\Gamma$ in Ba(Fe$_{1-x}$Co$_{x}$)$_{2}$As$_{2}$. Data from other publications are shown for comparison. $\Gamma_{c}$ is the critical scattering rate that corresponds to $T_{c}$ = 0. Dependence of $\frac{\Delta C}{\gamma_{n}T_{c}}$ (b) and the residual density of states $\gamma_{r}$ (c) on $T_{c}$. Open and closed symbols represent under- and overdoped concentrations, respectively. Magenta lines are fit to the data for a dirty $s\pm$ superconducting state (with $N_{1}(0)=N_{2}(0)=N(0)$, $\Delta_{1}=-\Delta_{2}=\Delta$) in the intermediate impurity-scattering strength {\it i.e.} for $\epsilon$ = 0.46.}  
\end{center}
\end{figure}

Unlike previous calculations using transport data to estimate the scattering rate,~\cite{Kirshenbaum12} our model accounts for the changes of $\Gamma$ with both $n_{i}$ and $N_{0}$, which are deduced from our experimental values of $\gamma_{n}$ shown in Fig.~\ref{Fig3}b. The effect of a chemical substitution in pnictides is quite clearly not describable solely in terms of a potential scatterer, but the impurity may dope the system or cause other electronic structure changes, which influence the pairing strength. In Fig.~\ref{Fig10}a, we show $T_{c}$ as function of $\Gamma/\Gamma_{c}$, where $\Gamma_{c}$ is the critical scattering rate for which $T_{c}$ = 0 and which corresponds to $x$ = 0.13 and $\gamma_{n}$ = 16.3 mJ mol$^{-1}$K$^{-2}$. The data are fitted with the Abrikosov-Gor'kov relation,~\cite{AG}
\begin{equation}\label{AGlaw}
-\ln \left(\frac{T_{c}}{T_{c0}}\right)=\psi \left(\frac{1}{2}+\frac{\Gamma}{2\pi k_{B}T_{c}}\right)-\psi \left(\frac{1}{2}\right),
\end{equation}
where $\psi \left(x\right)$ is the digamma function. We find that the fit reproduces the data very well for the overdoped concentrations ($x>0.063$), leading to $T_{c0}$ = 34 K for $\Gamma$ = 0. Our model however fails for $x<0.063$, because a finite magnetic order parameter is not taken into account in our model. The counterintuitive sharp increase of $T_{c}$ with increasing scattering rate in the underdoped samples was actually predicted theoretically in Refs ~\onlinecite{Vavilov11,Fernandes12}. Theoretically, this rise is expected in any density-wave superconductor, for both $s_{++}$ and $s\pm$ states,~\cite{Machida82,Gabovich83,Fernandes12} and is due to a stronger sensitivity of SDW or CDW to disorder than superconductivity. This is the reason why $T_{c}$ increases with $\Gamma$, and this is potentially a direction for future studies.     
As shown in Figs ~\ref{Fig3}b and ~\ref{Fig6}b, the residual density of states $\gamma_{r}$ is zero only at the optimal concentration $x=0.06$ which bounds the onset of gaplessness in Ba(Fe$_{1-x}$Co$_{x}$)$_{2}$As$_{2}$. It corresponds to a scattering rate of $\Gamma_{g}/{\Gamma_{c}}\approx 0.34$ (see Fig.~\ref{Fig10}a), which is only related to the strength of the impurity-scattering potential $\epsilon$ via,~\cite{Chaba72}
\begin{equation}\label{gapless}
\frac{\Gamma_{g}}{\Gamma_{c}}=2\epsilon^{2}\exp\left[\frac{\pi\epsilon^{2}}{2(1+\epsilon)}\right].
\end{equation}
Solving Eq.~\ref{gapless} leads to $\epsilon$ = 0.46 (or equivalently to $\delta$ = 63$^{\circ}$) indicating that scattering in Ba(Fe$_{1-x}$Co$_{x}$)$_{2}$As$_{2}$ is of intermediate strength. With these values of T$_{c0}$ and $\epsilon$, the dependence of $\frac{\Delta C}{\gamma_{n}T_{c}}$ and $\gamma_{r}/\gamma_{n}$ on T$_{c}$ can be calculated using the analytical expressions given in Ref.~\onlinecite{Chaba72} and are compared to our data in Figs ~\ref{Fig10}b and ~\ref{Fig10}c, respectively.   
We find that our model accurately reproduces the evolution of the heat-capacity jump with $T_{c}$ and yields $\left(\Delta C/\gamma_{n}T_{c}\right)_{0}$ = 2.25 for $\Gamma$ = 0. However, the agreement for $\gamma_{r}/\gamma_{n}$ is correct only at the onset of gaplessness near $T_{c}$ = 25 K. The change of curvature observed near $T_{c}$ = 10 K in the experimental data probably results from the existence of at least one additional smaller energy gap ($\Delta_{3}<\Delta$), which is not accounted for by our model. As mentioned earlier in Section~\ref{gapamp}, the larger gap mainly determines the jump at $T_{c}$, while the smaller ones govern the $T\rightarrow0$ behavior. This explains why our model works well for $\Delta C/\gamma_{n}T_{c}$ but not for $\gamma_{r}/\gamma_{n}$. Thus, a full treatment~\cite{Bang09,Efremov11} (including additional bands and probably strong-coupling corrections) in the $T$-matrix approximation is required to obtain full agreement with the experimental data. Nevertheless, our simplified model clearly shows that in-plane substitution of Fe by Co is strongly pair-breaking with an intermediate impurity-potential strength, in strong contrast to out-of-plane K substitution, which is not pair breaking. Furthermore, the "impurity-free" $T_{c}$ value obtained in this work, $T_{c0}$ = 34 K, is more realistic than the room-temperature value obtained in the Born limit~\cite{Kogan10} and is close to the optimal value found in Ba$_{1-x}$K$_{x}$Fe$_{2}$As$_{2}$ ($T_{c}$ = 38 K) and Ba$_{1-x}$Na$_{x}$Fe$_{2}$As$_{2}$ ($T_{c}$ = 34 K),~\cite{Wang16} which are essentially insensitive to potential scattering, as illustrated by the absence of gapless excitations in these compounds.

\section{Conclusion}

We have thoroughly explored the normal- and superconducting-state properties of Ba$_{1-x}$K$_{x}$Fe$_{2}$As$_{2}$, RbFe$_{2}$As$_{2}$ and CsFe$_{2}$As$_{2}$ both experimentally and theoretically. In the normal state, we find clear evidence of substantial correlations that are strongly enhanced with hole doping and with the isovalent substitution K $\rightarrow$ Rb $\rightarrow$ Cs. The strong differentiation of the mass enhancement among the different bands observed by quantum-oscillation experiments explain the prominent coherence-incoherence crossover observed for all these compounds. These results are well reproduced by DFT + SS calculations confirming that these materials are effectively Hund metals in which sizable Hund's coupling, orbital selectivity and doping are the key parameters for tuning the correlations. These systems behave as expected in the vicinity of a Mott insulator which could, in principle, be reached in an orbital-selective fashion by further hole doping.

In the superconducting state of Ba$_{1-x}$K$_{x}$Fe$_{2}$As$_{2}$, strong multiband features are clearly observed in the heat capacity, and no evidence for nodes is found, ruling out a doping-induced change of symmetry of the superconducting ground state. Thus, the symmetry remains $s$-wave. The system Ba$_{1-x}$K$_{x}$Fe$_{2}$As$_{2}$ is insensitive to out-of-plane disorder introduced by K doping, and the phase diagram is governed solely by the changes of the electronic structure. We attribute the suppression of T$_{c}$ and the larger energy gap in the range $0.4<x<1.0$ to the vanishing of the dominant interband electron-pair scattering caused by the disappearance of the electron band, which is consistent with a spin-fluctuation mechanism. Conversely, we argue that in-plane disorder is strongly detrimental to superconductivity in the Ba(Fe$_{1-x}$Co$_{x}$)$_{2}$As$_{2}$ series and that this scattering is of intermediate strength. In this case, pair-breaking is the primary reason for the suppression of superconductivity and changes of the electronic structure play only a minor role. Pair breaking is also shown to account for the 'pseudo' scaling of Bud'ko, Ni and Canfield which only holds for the electron-doped systems.


\begin{acknowledgments}
We thank Steffen Backes, Jacques Flouquet, Jean-Pascal Brison, Kai Grube, Felix Eilers, and Peter Hirschfeld for fruitful discussion.    
\end{acknowledgments}

\bibliography{biblio1}

\end{document}